 \documentclass[preprint]{aastex}

\shorttitle{White Dwarf Cooling Age of NGC 2099}
\shortauthors{Kalirai, J. S. et al.}

\begin{document}


\title{The CFHT Open Star Cluster Survey. III. The White Dwarf
Cooling Age of the Rich Open Star Cluster NGC 2099 (M37)}


\author{Jasonjot Singh Kalirai\altaffilmark{1}}
%\affil{Physics \& Astronomy Department, 6224 Agricultural Road,
%University of British Columbia, Vancouver, BC V6T-1Z1}
%\email{jkalirai@physics.ubc.ca}

%\author{Harvey B. Richer, Gregory G. Fahlman,
%Francesca D'Antona}

\author{Paolo Ventura \altaffilmark{2}}
%\affil{{Osservatorio Astronomico di Roma}
\author{Harvey B. Richer\altaffilmark{1}}
%\affil{University of British Columbia}
\author{Gregory G. Fahlman \altaffilmark{1,3}}
%\affil{Canada-France-Hawaii Telescope}
\author{Patrick R. Durrell \altaffilmark{4}}
%\affil{Penn. State University}
\author{Francesca D'Antona \altaffilmark{2}}
%\affil{Osservatorio Astronomico di Roma}

\and

\author{Gianni Marconi \altaffilmark{5}}
%\affil{ESO}


\altaffiltext{1}{Department of Physics \& Astronomy, 6224
Agricultural Road, University of British Columbia, Vancouver, BC
V6T 1Z1; jkalirai@physics.ubc.ca, richer@astro.ubc.ca}
\altaffiltext{2}{Osservatorio Astronomico di Roma, Sede di
Monteporzio Catone, Via Frascati 33, I-00040, Italy;
paolo@coma.mporzio.astro.it, dantona@coma.mporzio.astro.it}
\altaffiltext{3}{Canada-France-Hawaii Telescope Corporation, P.O.
Box 1597, Kamuela, Hawaii 96743, USA; fahlman@cfht.hawaii.edu}
\altaffiltext{4}{Department of Astronomy \& Astrophysics,
Pennsylvania State University, 525 Davey Lab, University Park, PA,
16802, USA; pdurrell@astro.psu.edu} \altaffiltext{5}{European
Southern Observatory, ESO-Chile, Casilla 19001, Santiago, Chile;
gmarconi@eso.org}


\begin{abstract}
We present deep CCD photometry of the very rich, intermediate aged
(similar to the Hyades) open star cluster NGC 2099 (M37).
The \it V, B$\rm-$V \normalfont color-magnitude diagram (CMD) for
the cluster shows an extremely well populated and very tightly
constrained main-sequence extending over 12 magnitudes from the
turn-off. Both a well defined main-sequence turn-off and a red
giant population are also clearly evident. The CFH12K photometry
for this cluster is faint enough (V $\sim$ 24.5) to detect the
remnants of the most massive progenitor cluster stars under the
Type I SNe limit. Therefore, the CMD of the cluster also exhibits
a well defined white dwarf `clump' caused by the decreased rate of
cooling of these stars as they age, and a subsequent gap with very
few objects. By using source classification to eliminate
faint-blue galaxies and a statistical subtraction technique to
eliminate foreground/background objects, we have determined the
age of the star cluster from the termination point (M$_{\rm V}$ =
11.95 $\pm$ 0.30) in the white dwarf luminosity function. The
white dwarf cooling age is found to be 566 $\pm \ ^{154}_{176}$
Myrs from comparisons with white dwarf cooling models and is in
excellent agreement with the main-sequence turn-off isochrone age
(520 Myrs). After carefully accounting for the uncertainties in
the white dwarf limiting magnitude, we show that the cooling age
confirms that models including convective core overshooting are
preferred for young-intermediate aged clusters. This is
particularly important in the case of NGC 2099 as the age is
similar to that of the Hyades cluster, for which current models
have difficulties in reproducing the details of the main-sequence
turn-off.  We also derive the reddening (E(B$\rm-$V) = 0.21 $\pm$
0.03) and distance ((m$\rm-$M)$_{\rm V}$ = 11.55 $\pm$ 0.13) to
NGC 2099 by matching main-sequence features in the cluster to a
new fiducial main-sequence for the Hyades, after correcting for
small metallicity differences. As a continuing part of the goals
of the CFHT Open Star Cluster Survey to better understand
dynamical processes of open clusters, we also fit a King model to
the cluster density distribution and investigate the cluster
main-sequence luminosity and mass functions in increasing
concentric annuli. We find some evidence for mass segregation
within the boundary of NGC 2099 as expected given the cluster's
age relative to the dynamical age. The present global mass
function for the cluster is found to be shallower than a Salpeter
IMF.

--------------------------------------

\end{abstract}


\keywords{color-magnitude diagrams -- open clusters and
associations: individual (NGC 2099) -- stars: evolution -- stars: luminosity function,
mass function -- white dwarfs}


\section{Introduction} \label{intro}

    White dwarfs represent the end point in the life of low
mass stars (M $\lesssim$ 7 M$_\odot$).  These burnt out stellar
cinders have extinguished their nuclear fuel and are now radiating
away any remaining stored thermal energy.  Younger than an age of
about 8 Gyrs, white dwarfs will simply cool and fade, becoming
dimmer and redder as time passes.  White dwarfs typically have 10
$\lesssim$ M$_{\rm V}$ $\lesssim$ 17.5 (depending on the age of
the star), making them difficult to detect at moderate distances.
This primary deterrent has prevented the establishment of
observational constraints on certain theoretical aspects of white
dwarfs. For example, constraints on both the initial-final mass
relationship for white dwarfs and the upper mass limit for white
dwarf production are very limited
\citep{weidemann1,weidemann3,reimerskoester7}.  A progenitor-white
dwarf initial-final mass relationship is particularly important
for Galactic chemical evolutionary studies as it would provide
important information on the amount of stellar mass loss in post
main-sequence evolutionary stages.

    Star clusters represent at least a partial solution to the
difficulties in studying white dwarfs and they provide an
excellent environment in which we can learn more about the
properties of these objects. The assumed single-burst star
formation event that creates a star cluster produces stars with a
spectrum of masses yet similar metallicity, age and (most
importantly) distance. Since the evolution of stars is driven
primarily by their initial mass, examining different populations
of stars in a cluster can give us a snapshot of the life stages of
a single star: turn-off, red giant, planetary nebula, white dwarf,
etc. \citep[see][]{renzini}. By applying a statistical argument to
objects in different locations of a cluster CMD, we can infer
information about the properties of objects in different stellar
evolutionary stages, such as white dwarfs.  Studying white dwarfs
in practice, however, is not this simple.  The richest star
clusters, the globular clusters, contain large numbers of white
dwarfs but these clusters are also very old. Therefore the
limiting magnitude of the coolest white dwarfs in these clusters
occurs at magnitudes that are too faint for ground based
observations. For example, the limiting white dwarf magnitude for
a 12 Gyr cluster occurs at M$_{\rm V}$ = 17.5 \citep{richer1}. The
only program currently underway to establish the white dwarf
cooling age of a globular star cluster is a deep HST study of M4,
the nearest globular cluster \citep{richer3}.  Success in this study may
be possible because it takes advantage of new models that predict radically
different behavior in the emergent spectra of cool white dwarfs
with ages $\gtrsim$ 8 Gyrs or T$_{\rm eff}$ $\lesssim$ 4000 K
\citep{hansen,saumon}.

An alternative method of studying white dwarfs is to utilize the
much younger and more metal-rich open star clusters.  These systems 
are found over a very broad range of parameters such as distance, 
age, metallicity and richness. However, there are also difficulties 
in using open star clusters to study white 
dwarfs.  First, the clusters are confined to the disk of the Milky
Way so background contamination is large. For poorly populated
clusters, this makes it very difficult to distinguish the cluster
population from the background.  The interpretation of the
photometry for many open clusters is also difficult due to heavy
extinction and reddening of star light. Second, most open clusters
are large in angular extent ($\sim$20$'$), so there are many
background galaxies in the field of view.  Some of these will
appear as faint, blue objects, thereby mimicking white dwarfs.
Thirdly, because open clusters are confined to the disk, tidal
interactions from giant molecular clouds work to dissipate the
clusters as they orbit the Galaxy. \cite{wielen} shows that a
moderately rich open cluster in the solar neighborhood consisting
of 500 stars ($\sim$250 M$_\odot$) will dissolve in a timescale of
about 2 $\times$ 10$^{8}$ years.

What we require to address the properties of white dwarf stars are
very rich, yet intermediate aged open star clusters with little
reddening.  The clusters must be old enough so that a significant
number of white dwarfs have formed, yet young enough so that the
end of the white dwarf cooling sequence occurs at an observable
magnitude.  Although NGC 2099 has a low Galactic latitude
(b$^{II}$ = 3.1$^{\rm o}$) and consequently a moderately high
reddening value (E(B$\rm-$V) = 0.21), it is an excellent cluster
for white dwarf studies because of its extremely rich stellar
population (our photometry detects 2620 stars) and its
intermediate age ($\sim$500 Myrs).

    NGC 2099 was established as a relatively rich and
large star cluster very early on and thus has been the focus of
many studies, for example: photometrically by
\cite{vonzeipel,becker1,hoag}; spectroscopically by
\cite{zug,lindblad}; and astrometrically by
\cite{nordlund,giebeler,lindblad,upgren} and \cite{jeffreys}. The
first significant photometric study that shed light on cluster
parameters was that of \cite{west}. West used photographic plates
to establish a CMD for 930 member stars brighter than V = 17.7.
The photometric spread in the cluster main-sequence from this
study is quite large and consequently, there are large
uncertainties in the results. That UBV study concluded with an
estimated cluster age of 220 Myrs, distance modulus of
(m$\rm-$M)$_{\rm V}$ = 11.6, and a constant reddening value of
E(B$\rm-$V) = 0.27.  The first significant dynamical study of the
stellar density distribution around the area of NGC 2099 was
based on three-color Schmidt plate photometry in 1975
\citep{becker2}. The latest study of the cluster used radial
velocity and UBV photoelectric observations to establish a much
cleaner main-sequence based on cluster membership
\citep{mermilliod}.  The results of this study concluded with
(m$\rm-$M)$_{\rm V}$ = 11.5, E(B$\rm-$V) = 0.29, age = 450 Myrs
(comparison to theoretical isochrones from Schaller et al. (1992)
and Bertelli et al. (1994)) and Z = 0.020.  The CMD of this study
exhibits less than 100 objects and has a limiting magnitude
brighter than V = 17. The quality of the data in the present study
is unprecedented when compared with these previous efforts (see
Figure \ref{comparisonfig} for a comparison of CMDs).

    In the next section we discuss the instrumentation that was
used to collect this data and the photometric reduction
procedures. We then present the cluster CMD which exhibits a very
tight main-sequence stretching over 12 magnitudes in the \it V,
B$\rm-$V \normalfont plane, consisting of more than 2600 stars. We
derive the reddening and distance modulus to NGC 2099 by comparing
the un-evolved main-sequence to a new fiducial for the Hyades star
cluster \citep{deBruijne}.  We also draw comparisons, both on the
observational and theoretical planes, to the Hyades cluster which
has a similar age to NGC 2099.  This age estimate is determined by
fitting the main-sequence turn-off and the luminosity of the red giant 
clump with up-to-date theoretical 
stellar evolutionary sequences, calculated especially for the CFHT
Open Star Cluster Survey (Kalirai et al. 2001a, hereafter Paper
I). An analysis of dynamical effects in the cluster is discussed
in \S\S \ref{extent} and \ref{lumfunc}.  The cluster mass function 
is derived in \S \ref{massfunc} and the corresponding total mass of 
the cluster is corrected for those stars between our limiting 
magnitude and the hydrogen burning limit in \S \ref{faintmsstars}.  
Finally, we conclude by presenting studies of potential white 
dwarf stars in NGC 2099 and determining the white dwarf cooling 
age of the cluster (\S \ref{wds}).

\section{Observations and Reductions}\label{observations}

    Most of the observational data for NGC 2099 was obtained during
the second night of a three night observing run in 1999 October,
using the CFH12K mosaic CCD on the Canada-France-Hawaii Telescope.
This CCD contains 12, 2048 $\times$ 4096 pixel (0$\farcs$206)
chips that project to an area of 42$'$ $\times$ 
28$'$ on the sky.  This is larger than the apparent cluster 
radius of $\sim$13$\farcm$9 (see \S \ref{extent}). We obtained three 
300-second images in both B and V, as well as single 50 and 10-second 
images in each of the B, V and R filters.  At a later date, we 
also acquired short 0.5-second frames to obtain unsaturated images
of the bright stars in the cluster (see Table 1). Blank field
images are not necessary as the outer regions of the mosaic can be
used to correct for field star and galaxy contamination. Most
images were obtained while the sky was photometric with sub-arcsecond
seeing conditions (see Table 1).  A color image created from the
deep V and B images is shown in Figure \ref{newmosaic}.

The data was processed (flat-field, bias and dark corrected) and
montaged using the FITS Large Images Processing Software (FLIPS)
(Cuillandre 2001) as described in \S 3 of Paper I. In the final
processed images, we find the flat-fielding to be good to
$\sim$0.5\% in V and $\sim$0.7\% in B, averaged over 11 
${\sq^{\prime\prime}}$ patches.  We reduced the data using a 
preliminary version of the 
new TERAPIX photometry routine PSFex (Point Spread Function
Extractor) (E. Bertin  2000, private communication).  We used a
separate, variable PSF for each CCD in the mosaic.  The mean
errors in the photometry were 0.015 mag at V = 22, 0.032 mag at V
= 23 and 0.075 at V = 24.  A statistical error plot is shown in
Figure \ref{errfig}. Further information on PSFex can be found in
\S 4 of Paper I. Numerous photometric calibration frames of
Landolt fields SA-92 and SA-95 \citep{landolt} were used to
calibrate the NGC 2099 data (see Tables 2 and 3 in Paper I).
Calibration methods are discussed in \S\S 5.1 and 5.2 of the same
paper.  The photometric uncertainty in the zero points for the
standard stars during night 2 were measured to be $\sim$0.021 in V
and $\sim$0.025 in B. The extinction in magnitudes per air-mass
was determined to be 0.084 $\pm$ 0.012 in V and 0.184 $\pm$ 0.008
in B, both close to the standard CFHT values of 0.10 and 0.17
respectively and similar to those on the first night of the run
(0.088 $\pm$ 0.010 in V and 0.165 $\pm$ 0.005 in B). The color
terms were averaged over the three night observing run and found
to be  in agreement with CFHT estimations in the V filter and
slightly lower than estimations for the B filter.

\section{The NGC 2099 Color-Magnitude Diagram}\label{cmd}

In Figure \ref{cmds2099} we present CMDs for both the cluster and
a background field after applying a 0.50 stellarity cut from
SExtractor \citep{bertin1} to remove obvious faint galaxies (see
\S \ref{stellarity}).  The cluster field in this Figure has been
scaled down in area to be consistent with the area represented in
the background field (i.e. each object in the background field
represents 1.37 cluster field objects).  This has been done by
eliminating a ring of objects centered close to the half-mass
radius (4.5 pc) of the cluster (a full CMD consisting of the entire cluster
population is presented later, Figure \ref{comparisonfig}).  The
cluster CMD shows a magnificent tightly constrained main-sequence
extending over 12 magnitudes from the cluster turn-off V $\sim$ 12
(M$_{\rm V}$ = 0.45) to V $\sim$ 23.5 (M$_{\rm V}$ = 11.95). A
tight red giant clump is seen consisting of 20 stars (see Figure
\ref{comparisonfig} for all stars). There is also a significant
background disk star population below the main-sequence in the
cluster CMD and in the background field.  This distribution arises
due to the low Galactic latitude of the cluster and clearly
overlaps significantly with the faintest cluster members on the
CMD.  In fact, the feature in the background field at 20 $\leq$ 
V $\leq$ 23.5 and B$\rm-$V $\sim$ 1.7 (which can also be seen in the 
NGC 6819 data set) may in part be faint cluster red stars located between 
our apparent cluster radius (the limit to which we can
detect cluster stars) and the tidal radius (see \S \ref{extent}).

The cluster CMD in Figure \ref{cmds2099} also shows several faint,
blue objects that will be discussed in detail in \S \ref{wds}.
Although there is field star and background galaxy contamination,
the density of objects is clearly greater in the cluster CMD
indicating that a significant number of these objects are cluster
white dwarf members (see \S \ref{wdcontamination} for statistical 
arguments involving the subtraction of field stars and removal of 
galaxies). These stars have quickly evolved
($\sim$10$^{4}\rm-$10$^{5}$ years) from the red giant phase to the
white dwarf phase and are now slowly piling up on the CMD at a
magnitude that can be used to independently infer the cluster age.
The NGC 2099 CMD clearly shows the faintest objects in this
`clump' to be at V = 23.5, after which point there is a
significant gap signifying that we have detected the termination
point in the white dwarf cooling sequence (the complete CMD in
Figure \ref{comparisonfig} shows this better).  We also note that
the photometry for these bluer objects is fainter than for the
main-sequence by over 1 magnitude.  

The cluster CMD (from Figure \ref{comparisonfig}) also shows several 
(5) very blue and bright 
objects, some of which may in fact be stars that have shed their
outer layers during the planetary nebula stage and are now cooling to
become white dwarfs.  We do not expect to detect many, if any,
objects at this stellar evolutionary stage because the stars spend
a very small amount of time during this phase.  A crude estimate
of the expected number of such objects can be determined by
comparing the evolutionary time ($\lesssim$10$^{5}$ years) of
these stars to the numbers (20) and timescales (100 Myrs) of red 
giants in the clump of NGC 2099. Although the theoretically expected
number is less than one, three of the objects in the CMD are centrally
concentrated in the cluster, not image defects, and have a high
stellarity index. Spectroscopic confirmation is needed to
determine whether any of these objects are cluster members and
potentially among the most interesting stars in NGC 2099.  If
confirmed, these objects would provide crucial observational
constraints on the phases of evolution from the planetary nebulae
stage to the white dwarf stage \citep{iben} and on neutrino
cooling of pre-white dwarf stars \citep{o'brien}.

\subsection{Stellarity} \label{stellarity}

    We use source classification to distinguish between stars
and galaxies at faint magnitudes where the signal-to-noise ratio
is low. The star/galaxy cut will affect both evolutionary tests of
stars that become white dwarfs and the luminosity function for
these stars.  We use SExtractor to assign a stellarity index to
all objects on all CCDs in our data. This stellarity index is
determined through a robust procedure that uses a neural network
approach as described in \cite{bertin1}.

    Figure \ref{stellfig} shows the variation of this stellarity
index with magnitude.  There is a clear group of both resolved
sources (stellarity = 0) and unresolved (stellarity = 1).
Additionally, we have visually inspected and determined that most
of the objects in the `clump' between 23 $\leq$ B $\leq$ 24.5 and
0.75 $\leq$ stellarity B $\leq$ 0.95 are most likely stars.  The
determination is more difficult for objects with lower stellarity.
Previous experience (Kalirai et al. 2001b, hereafter Paper II) has
shown that a 0.50 stellarity cut may not be stringent enough
whereas a 0.75 stellarity cut may be eliminating too many stellar
sources. For now, we will employ a 0.50 stellarity cut (in all
Figures) so as to not remove any truly stellar objects.
This cut will be justified in \S \ref{wdcontamination}.

\subsection{Cluster Metallicity}\label{metallicity}

An ideal testing of theoretical stellar isochrones by using
observational CMDs requires prior knowledge of the cluster
metallicity, reddening and distance.  Unfortunately, prior NGC
2099 data have not set tight constraints on any of these.
Although the published literature does not show any detailed
spectroscopic abundance studies of the cluster,  Lynga's Fifth
Catalogue of Cluster Parameters \citep[see e.g.,][]{janes} lists the
cluster metallicity to be [Fe/H] = 0.09 or Z $\sim$ 0.020.
Additionally, the study of \cite{mermilliod} experimented with
different models (Schaller et al. 1992 and Bertelli et al. 1994)
and examined the colors of the red giant clump in NGC 2099 also
concluding with a best fit Z = 0.020 isochrone.  Since there have
been no studies contradicting this metallicity, we adopt the Z =
0.020 value as the most likely cluster metallicity, but will also
consider a Z = 0.025 (Hyades-like \citep{perryman}) value (see 
\S \ref{obstheory}).

\subsection{Cluster Reddening and Distance -- Main-Sequence Fitting
with the Hyades}\label{parameters}

The only previous model-independent reddening estimate of NGC 2099
comes from the photographic work of \cite{west}, E(B$\rm-$V) =
0.27.  The main sequence of these data contains a
large scatter.  The color-color plots also suffer from large
uncertainties (photometric spread $\sim$0.5 magnitudes in
B$\rm-$V).  The study of \cite{mermilliod} compared their
photoelectric data to stellar isochrones and estimated the 
reddening to be E(B$\rm-$V) = 0.29.  The advantages of
determining a reddening value from the present data over the
previous works include the higher stellar density, deeper CMD and
lower uncertainties in the photometry.  See Figure
\ref{comparisonfig} for a comparison of the present CMD with these
studies and Figure \ref{matchfig} for a residual star-by-star comparison 
with the photoelectric data.  Unfortunately, the R filter photometry 
for NGC 2099 was 
unable to provide reddening constraints as the slope of the
reddening vector (E(V$\rm-$R)/E(B$\rm-$V) = 0.78) is too similar
to the observed sequence in the color-color plot.  An accurate UBV
or BVI analysis is desirable for the cluster.

An alternative method of determining the reddening of a star
cluster was demonstrated in the analysis of M4 \citep{richer3},
where a set of subdwarf stars were fit to the observed
main-sequence to simultaneously solve for both the reddening and
apparent distance modulus.  Due to the degeneracy between these
two quantities, the results established more stringent ranges on
the reddening than for the apparent distance modulus.  We adopt a
similar technique by using the well established Hyades
main sequence to fit NGC 2099.  This method is ideal for the
present comparisons due to the long main sequences of the two
clusters (stars with mass below 2.6 M$_\odot$ in NGC 2099 have not
yet evolved off the main-sequence), accurate photometry (several
observable features are present in both main-sequences), similar
metallicity of the clusters (Z = 0.024 for Hyades) and similar
age for the clusters (see \S \ref{obstheory}).

The new fiducial main-sequence of the Hyades open star cluster is
based on a new analysis of Hipparcos individual parallaxes by
\cite{deBruijne}.  The fiducial was calculated by binning known
cluster members at 0.5 magnitude intervals from M$_{\rm V}$ = 1.5
to M$_{\rm V}$ = 9 with a small color cut to remove outliers.  
Binaries were removed, however we include the brightest binary 
component which defines the turn-off of the 
cluster (star $\theta^{2}\tau_{A}$).  We first adjust the NGC 2099
CMD by +0.03 in B$\rm-$V to account for the $\Delta$Z $\sim$ 0.005
metallicity difference in the clusters.  We then compare
observable features in our tight main-sequence, such as the `kink'
in the main-sequence at (B$\rm-$V)$_{\rm o} \sim$ 0.4 caused by
changes from convective envelope models to radiative envelopes, as
well as the general shape and slope of the main-sequence to the
Hyades fiducial (E(B$\rm-$V) = 0.003 $\pm$ 0.002, mean 
(m$\rm-$M)$_{\rm V}$ = 3.27) to solve for the best 
combination of reddening and apparent distance modulus.  The 
results point towards a reddening of E(B$\rm-$V) = 0.23 $\pm$ 0.03 
and an apparent distance modulus of (m$\rm-$M)$_{\rm V}$ = 11.65 
$\pm$ 0.13, where the uncertainty in the latter value reflects the 
range of main-sequence fitting 
distance moduli obtained by using reddening values at the
extremes of the $\pm$0.03 uncertainty. Figure \ref{hyadesfig}
illustrates the remarkable agreement between the detailed features
of the two main-sequences for these values. Additional reddening
constraints are also provided by comparing the colors of the NGC
2099 red giant clump to the Hyades clump (which is reproduced well
by our current theoretical models).  The small age difference
between the two clusters (see \S \ref{obstheory}) would affect the
luminosity of the clump stars significantly, but only slightly
affect the color. On this metallicity corrected plane, the clump
of NGC 2099 can not be redder than the Hyades (would imply Z 
$\gtrsim$ 0.024), however for a reddening value
E(B$\rm-$V) $\lesssim$ 0.18, this condition is violated and
therefore such a low reddening can be ruled out.

    Given the uncertainties in the reddening, distance modulus, 
metallicity and model colors (discussed in \S \ref{obstheory}) we 
incrementally massage the reddening by $\rm-$0.02 to allow a slightly 
better isochrone fit of the model to the un-evolved main-sequence of NGC
2099. Therefore, our best estimate of the reddening of NGC 2099 is
E(B$\rm-$V) = 0.21 $\pm$ 0.03, with a corresponding distance
modulus of (m$\rm-$M)$_{\rm V}$ = 11.55 $\pm$ 0.13. The
uncertainties in these values are not known well enough to
establish different upper and lower limits based on the adjustment
from the main-sequence fitting values. The higher limits
established earlier are ruled out by the models so to remain
consistent we will use the model-massaged values throughout. With 
the lower adopted reddening value, this distance modulus is in 
good agreement with the value of West ((m$\rm-$M)$_{\rm V}$ = 
11.64 $\pm$ 0.1) and Mermilliod ((m$\rm-$M)$_{\rm V}$ = 11.50). The 
agreement with the Hyades cluster using the new reddening and distance 
modulus is slightly worst than for the best main-sequence fitting 
values.  This is most likely due to the metallicity of the cluster not 
being exactly Z = 0.02.  The corresponding distance to NGC 2099, using 
A$_{\rm V}$ = 3.1E(B$\rm-$V), is 1513 $\pm \ ^{146}_{133}$ pc.  The error 
in the true distance modulus (10.90 $\pm$ 0.16) combines four individual
errors, and accounts for correlations between them: (1) a scale
factor translation ($\Delta(B-V)$ to $\Delta{V}$) from the
B$\rm-$V axis to the V axis to account for the reddening
uncertainty, (2) an estimated main-sequence fitting uncertainty at
a fixed reddening value, (3) the error in the extinction,
$\Delta_{A_{V}}$ = 3.1$\Delta_{E(B-V)}$ and (4) the color
uncertainty due to an estimated metallicity uncertainty,
$\Delta_{Z}$. We evaluate and provide a more detailed description
of these terms in \S \ref{uncertainty}.

\subsection{Theoretical Isochrones}\label{theory}

\subsubsection{Model Description}\label{models}

The theoretical models that we will use for this project have been
calculated especially for the CFHT Open Star Cluster Survey at the
Rome Observatory and are up-to-date in the input physics.  A
detailed description of the stellar evolution code that was used
to build the tracks can be found in \cite{ventura} (also see \S
5.2 of Paper II). We assume that the helium abundance scales with
the metallicity content according to the relation
$\Delta$Y/$\Delta$Z = 2. The models adopt convective
core-overshooting by means of an exponential decay of turbulent
velocity out of the formal convective borders as fixed by
Schwarzschild's criterion; this behavior of velocity is consistent
with approximate solutions of the Navier-Stokes equations
\citep{xiong}, and with the results of numerical simulations
\citep{freytag}.  A value of $\zeta$ = 0.03 for the free parameter
giving the e-folding distance of the exponential decay has been
adopted \citep{ventura}.  The convective flux has been evaluated
according to the Full Spectrum of Turbulence (FST) theory
prescriptions \citep{canuto2}. The theoretical isochrones were 
transformed into the observational plane by making use of 
the \cite{bessell} conversions.  The lower main sequence 
(M $\lesssim$ 0.7 M$_\odot$) has been calculated 
by adopting NextGen atmosphere models \citep{hauschildt}. Models 
based on a grey atmosphere approximation give almost identical 
results for larger masses.  For M $\lesssim$ 0.47 M$_\odot$ (or 
T$_{\rm eff}$ $\lesssim$ 3500 K) the transformations 
of \cite{hauschildt} in B$\rm-$V are not very reliable and so 
the faint end of the isochrones terminate at this mass.  However, 
the bolometric correction for cooler stars in the models can still 
be used to establish a mass-luminosity relationship extending to 
M $\sim$ 0.25 M$_\odot$ (this is used in \S \ref{massfunc}).

\subsubsection{Observation vs. Theory}\label{obstheory}

Comparing the observed NGC 2099 CMD with stellar isochrones can
not only provide an age for the cluster, but can also allow for
refinements in the models and a better understanding of the theory
of stellar evolution. Figure \ref{3isos} shows three panels with
two isochrones assuming extra mixing beyond the formal convective
borders, but for different ages, and one isochrone calculated
assuming no extra mixing.  We argue the best fit in Figure 
\ref{3isos} is in the left panel (520 Myrs - assuming core-overshooting) 
and for this combination of reddening, distance, metallicity and age, the
isochrone reproduces several features of the CMD nicely.  Several
comparisons can be drawn from the fit of the theory to the
observations.  First the lower main-sequence in NGC 2099 is
clearly redder than the isochrone.  This slope change in the
main-sequence, caused by the onset of H$_{\rm 2}$
dissociation-recombination in the stellar envelope
\citep{copeland} is not reproduced well by stellar models
employing grey atmosphere boundary conditions (see e.g., 
Castellani, Degl'Innocenti \& Prada Moroni 2001 for the lower 
main-sequence of the Hyades).  The non-grey boundary conditions 
employed in our models for the lower 
main-sequence (J. Montalban 2001, private communication) are in
better agreement with the general shape of the main-sequence, but
the models are still too blue by $\sim$0.05 mag in B$\rm-$V.  This
disagreement can be explained in part due to the use of the
\cite{bessell} colors at such low temperatures.  Photometric
spread in the data for faint stars may also partly contribute to
such an effect.  Also, the turn-off of NGC 2099 falls slightly
redder than the model, although the shape is consistent with the
data.  Despite considerable effort, we find it difficult to
reproduce the details of this feature in the NGC 2099 CMD with
current models (while maintaining a good fit to the luminosity of
the red giant clump). The problem here is very similar to that for
the Hyades cluster, for which several groups are unable to provide
a good fit to the turn-off. An extensive discussion concerning the
difficulty of fitting both the turn-off and the lower main
sequence region of the Hyades by adopting the same set of
color-temperature calibrations can be found in \cite{deBruijne}.
The fact that the NGC 2099 data is in excellent agreement with the
Hyades data (see Figure \ref{hyadesfig}), and models fail to
reproduce the turn-off feature in both clusters, will motivate a
deeper understanding of this feature of the CMD for intermediated
aged clusters. Conversely, both the color and luminosity of the
red giant clump are in excellent agreement with the data; the
latter provides the age.  The shape and slope of the main-sequence
(from M$_{\rm V}$ = 2.5 $\rm-$ 8.0) is also in excellent agreement
with this model.  For example, the `kink' in the 
main-sequence at $\sim$1.3 M$_\odot$ described earlier is modelled
perfectly.  The age estimate of 520 Myrs for NGC 2099 is also
consistent with Figure \ref{hyadesfig} which shows both the
peel-off of the main-sequence stars in the Hyades (age $\sim$ 600
Myrs with our models) and the luminosity of the clump stars to be
slightly fainter than that for the corresponding stars in NGC
2099.

Fitting exactly the location of the turn-off region of the cluster
requires a slightly higher age.  The middle panel of Figure
\ref{3isos} shows the fit for an age of 580 Myrs.  The
shape of the turn-off of the cluster, as well as the hook at the top
of the main-sequence caused by the contraction of stars that have
exhausted their hydrogen supply (very difficult to see), are
both modelled well.  However, a close examination of the luminosity of
the red giant clump shows a discrepency of $\sim$0.20 magnitudes.
Given the importance and relative simplicity of fitting the clump
luminosity rather than the turn-off, we believe this to be a
poorer fit than that presented in the left panel.

Due to the uncertainty in the reddening value established in
\S \ref{parameters}, we also note that we can not
completely rule out ages as low as 450 Myrs for NGC 2099 (the
value found by \cite{mermilliod}).  Using a reddening value
near our upper limit (E(B$\rm-$V) $\sim$ 0.25), we obtain a decent
fit to the luminosity of the clump stars, however, the subsequent
fit to the main-sequence for this reddening is poor.

We also examined the possibility of fitting the observed NGC 2099
stars with an isochrone of metallicity Z = 0.025.  For this, we
derived the reddening and distance modulus by comparing the NGC
2099 main-sequence with the Hyades fiducial with no metallicity
shift.  The corresponding best fit to the luminosity of the clump
is obtained for an age of $\sim$550 Myrs.  The NGC 2099 clump is
brighter and slightly redder than the Hyades clump, indicating a
younger age for NGC 2099 with respect to the Hyades.  The fit to
the slope of the main-sequence for this metallicity is clearly
worse than for a Z = 0.020 model.

As we will discuss later in \S \ref{wds}, one of the possible uses
of white dwarfs as chronometers is to check the consistency of the
turn-off and white dwarf ages.  If we fit the cluster CMD with
theoretical isochrones based on stellar models of the same
metallicity (Z = 0.020), but calculated assuming no extra mixing
beyond the formal convective borders, the best agreement between
the theory and observations is obtained for an age of
$\sim$300$\rm-$350 Myrs, using the same distance modulus and
reddening as in the overshooting case.  In this case, which is
based on reproducing the luminosity of the clump, we get a very
poor fit to the turn-off region, which is noticeably bluer than
the observed stars (see right panel of Figure \ref{3isos}). An age
of $\sim$400$\rm-$450 Myrs would lead to an acceptable fit of the
turn-off of the cluster, but in this case the theoretical
luminosity of the clump would be $\sim$0.4 magnitudes fainter than
the observed clump stars, a gross disagreement.

Finally, we tested to see what derived age would result for NGC 2099 from 
using the latest Padova group isochrones \citep{girardi}.  For their 
models with Z = 0.019 and Y = 0.273 (close to our values), we find a good 
fit to the general main-sequence using our derived cluster reddening and 
distance modulus.  The best age however, is slightly younger (480 Myrs).  
This is most likely due to a slightly lower metallicity in their models and 
a different treatment of convection (a scaled overshooting parameter based 
on the turn-off mass).  The theoretical clump in the Padova models is 
redder and fainter than the data (which again results from a different 
description of convection).

The above comparisons suggest that a certain amount of
overshooting from the border of the convective core is required
and that a solar metallicity (Z = 0.020) model of age 520 Myrs
provides the best fit to the observed NGC 2099 CMD (based on the 
luminosity of the clump).  However, a slightly larger age provides 
a better fit to the turn-off of the cluster.  Models without 
convective core overshooting do not reproduce the observed 
data well.  Ultimately it is the white dwarf cooling age 
of the cluster that will be the decisive discriminant (see 
\S \ref{wdcoolingage}).

\section{Star Counts}\label{starcounts}

\subsection{Center of the Cluster}\label{center}

    Star counts in concentric annuli around a cluster require a
determination of the center of the cluster.  To estimate this, we
produced a histogram of star counts in thin strips across the
mosaic.  The center, in each of the x and y directions, was
estimated from the center of the profile.  The results give 
coordinates $\alpha_{J2000} = 05^{h}52^{m}17.6^{s}, \delta _{J2000} =
+32^{\rm o}32'08''$ which however are limited in accuracy to about 
30$''$.  The uncertainty arises because of the large 
amount of scatter in the resulting plots (small number statistics).

\subsection{Radial Density Distribution}\label{extent}

Figure \ref{kingfig} shows the number of stars in increasing
concentric annuli around NGC 2099, normalized by the area of each
annulus.  The annuli used are each 2$'$ in width.  The background
field star population has been removed by scaling the counts in
the background field (taken from the outer four CCDs surrounding
the cluster -- 14.4 stars/${\sq^{\prime}}$) to the area of each annulus.
To avoid significant incompleteness biases, only stars with V
$\leq$ 22 have been used. Also shown is a single mass King model
\citep{king,king2} which is found to fit the data well.  We fit
the model to the data by varying the concentration parameter
(Log($r_{t}$/$r_{c}$)) until the best fit is obtained. This allows
us to determine the core radius, $r_{c}$ = 2.3 pc, which strictly
speaking has no physical meaning.  The model then gives us the
tidal radius, $r_{t}$ = 20.4 pc, which is caused by tidal
influences of massive objects in the Milky Way (e.g., GMCs) that
remove the highest-velocity stars from the cluster as they venture
out to large distances from the center.  Finally, we compare the
tidal radius from the model to the expected tidal radius of the
cluster given the cluster's mass and it's position in the Galaxy 
\citep{binney},

%-----------------------------
\begin{equation}
r_{t} \sim (\frac{m}{3M})^{1/3}D. \label{tidalequation}
\end{equation}
\medskip
%-----------------------------

\noindent The mass of the cluster in this calculation (m = 2515
M$_\odot$) has been corrected by including those stars between our
faintest magnitude bins and the hydrogen burning limit (0.08
M$_\odot$) as described in \S \ref{faintmsstars}.  M is the mass
of the Galaxy within the cluster's orbit (1.02 $\times$ 10$^{\rm
11}$ - \cite{clemens}) and D is the Galactocentric distance of NGC
2099 ($\sim$10 kpc, almost directly towards the anti-center of the
Milky Way).  Solving equation \ref{tidalequation} for the tidal
radius of NGC 2099 gives $r_{t}$ = 20.2 pc, in excellent agreement 
with the model value.  

	The apparent cluster radius of NGC 2099 is only $\sim$6 pc 
and therefore some stars may have evaporated past this radius 
towards the outer parts of the cluster (this may in part be an explanation 
for the faint, red feature in the background CMD -- see Figure 
\ref{cmds2099}).  Such an effect would be important as it would 
bias the faint end luminosity function.  We use the King model 
to determine the total mass of NGC 2099 out to the tidal radius of 
the cluster.  This involves calculating the projected central density 
of the cluster as well as several model dependent parameters 
\citep[see][]{gunn}.  This value is found to be 2548 M$_\odot$ 
and is in excellent agreement with the value found by integrating 
the mass function to our limiting radius (2515 M$_\odot$ -- see 
\S\S \ref{massfunc} and \ref{faintmsstars}).  Further discussion 
of the dynamical state of the cluster can be found in \S 
\ref{masssegregation}.

\section{Luminosity Functions}\label{lumfunc}

    The rich stellar population of NGC 2099 motivates star count
studies to determine the cluster population, its dynamical state,
and investigate the luminosity and mass functions.  The shape of
the luminosity function for this cluster is very important to
establish observational constraints on the timescales and levels
at which we expect dynamical evolutionary effects in clusters to
take place. Dynamical effects in clusters are caused by
equipartition of energy between stars of different masses.
Although the concentration of stars for any mass will be highest
in the center of the cluster and decrease as a function of radius
from the center \citep{binney}, as the cluster ages we can expect
the density distribution to expand from the center.  The low mass
stars will gain energy in the interactions and have higher
velocity dispersions than the high-mass stars, which sink to the
center of the cluster.   The relaxation timescale is proportional 
to the number of crossings of a star 
across the cluster that are required for its velocity to change by
order of itself \citep{binney}.  Using the distance (1.5 kpc),
population size (see \S \ref{faintmsstars}), linear size (6.2 pc)
and mass (see \S \ref{faintmsstars}) of NGC 2099, we determine the
relaxation time (see equation \ref{eqnrelaxationtime}) to be
$\sim$300 Myrs, 

%-----------------------------
\begin{equation}
t_{relax} \sim t_{cross}\frac{N}{8lnN}. \label{eqnrelaxationtime}
\end{equation}
\medskip
%-----------------------------

\noindent Therefore, we can expect the cluster (age $\sim$
500 Myrs) to be relaxed and exhibit some dynamical effects such as 
an excess of lower mass stars in the outer regions of the cluster.
In fact, if old enough the cluster may even lose some stars due to
these and other external processes such as tidal interactions in
the disk of the Galaxy \citep{wielen}.

    We define the cluster stars by first creating a main-sequence
fiducial (clipping objects with (B$\rm-$V) $\geq$ 3.5$\sigma$ from
the mean) after isolating the main-sequence from the background
distribution (see Table 2).  We then use a clipping routine to create an
envelope around this fiducial based on the errors in the photometry
(envelope broadens out towards faint magnitudes).  The counting of
the stars is done within this envelope, for both the cluster CMD
and the background CMD, with the raw cluster luminosity function
coming from the difference between the counts in the two fields 
(after accounting for aerial differences - see next section).

\subsection{Incompleteness Corrections and Counting Uncertainties}
\label{incompcorr}

    Before we can interpret results from the star counts in NGC 2099 
we must correct our data for incompleteness in the number of detected 
objects.  This effect is typically negligible for brighter objects and
increases for fainter sources. We produce an artificial catalogue
of input stars for which we know the magnitudes and colors.  A
small number of these stars is added uniformly in several trials
(7) in proportion to their numbers in the raw cluster luminosity
function so as not to affect the crowding statistics of the field.
These stars are chosen along the same slope of the main-sequence,
and a separate sequence for the location of the white dwarfs.
After adding these stars to our data frames, we re-reduce the new
data in an identical manner to the original data in the cluster.
This involves running PSFex on the data and merging the output V
and B files.  We then count the number of stars per magnitude bin
that were recovered (in both V and B). This analysis is carried
out for 2 cluster CCDs and 2 background field CCDs in order to
make the analysis reliable. The cluster CCDs represent 2 of the
central chips of CFH12K, and provide statistics on a 14$'$
$\times$ 14$'$ region near and including the central core of NGC
2099. Each of the background field CCDs represent a 14$'$ $\times$
7$'$ area taken on opposite sides of the cluster.  The NGC 2099
data set is found to be 100\% complete down to V = 19 in the
background field and 100\% complete to V = 17 in the cluster
field. For main-sequence stars the correction at V = 22.5 in the
cluster field is 1.219 (i.e. N$_{\rm added}$/N$_{\rm recovered}$ =
1.219). The white dwarfs are more complete than the main-sequence
stars in almost all bins. This is expected as these stars are
brighter in the B band for a given V band magnitude than the
main-sequence stars. A summary of the incompleteness factors for
the 2 cluster and background field CCDs is given in Table 3.

    The errors in the incompleteness corrections are evaluated
using an identical analysis to that described by \cite{bolte} and
implemented in Paper II.  Namely, we assume that the counting
uncertainties are derived from a Poisson distribution and that the
artificial star count uncertainties are derived from a binomial
distribution.  Furthermore, the errors in the incompleteness and
the raw star counts are assumed to be uncorrelated.  The errors
from the cluster and background fields are added in quadrature.

    The completeness-corrected number of stars in the cluster can
now be determined in three steps. First, we multiply the cluster
field incompleteness correction by the number of stars in the
cluster in that magnitude range. Second, we multiply the background 
field incompleteness 
correction by the number of background field stars, and multiply
the result by 1.37 to account for the difference in aerial
coverage of the background. Finally, we subtract the two and
obtain the corrected star counts.  For the global star counts, we
use the inner 13$\farcm$9 for the cluster field and establish a
background field area from the outer portions of the outer CCDs.
The cluster density at this radius has dropped off significantly
from the central regions, however, in the absence of a strong
field star population, cluster members could have been detected at
greater radii.

The final corrected star counts are presented in Table 4. In this
table, the first row of each magnitude bin consists of raw counts
(cluster field $\rm-$ 1.37$\times$background field) whereas the
row underneath contains the incompleteness corrected numbers
(correction applied from Table 3). Also shown in parentheses are
the errors in these counts, as calculated from the analysis given
above.

\subsection{Analyzing the Cluster Luminosity
Function}\label{lumfuncanalysis}

    The global luminosity function for NGC 2099 is plotted in Figure
\ref{lumfuncfig}, where the dashed line represents the raw counts
and the solid line the incompleteness corrected counts.  The
global luminosity function is almost flat from V = 13.5 (M$_{\rm
V}$ $\sim$ 2) to V = 19.5 (M$_{\rm V}$ $\sim$ 8) and slowly rises
beyond this point.  This rise is due to the change of slope in the
mass-luminosity relation at M$_{V} \sim$ 8 for solar chemistry 
\citep[see e.g.,][]{dantona}, which can also be seen as a change in
slope in the cluster main-sequence (see \S \ref{obstheory} or CMD
figures).  Integrating the luminosity function and accounting for
red giants and white dwarfs provides a lower limit to the total
cluster population of $\sim$2600 stars (this value is corrected
for stars down to the H-burning limit in \S \ref{faintmsstars}).
This observed NGC 2099 stellar content makes it very similar to
that of NGC 6819 (2900 observed stars), and therefore one of the
richest open star clusters known.

    In Figure \ref{lumfuncfigcomp} we compare the NGC 2099
luminosity function to that for the Solar Neighborhood
\citep{binney2}, the Pleiades \citep{lee} and NGC 6819 (Paper
II).  We have normalized the luminosity functions to the number of
stars in NGC 2099 at the M$_{\rm V}$ = 5 bin.  The masses of the
stars in this bin are slightly less than solar and no stellar
evolutionary effects have yet taken place for any of the clusters.
The luminosity function for NGC 2099 exhibits a similar slope to
both the Pleiades and the Solar Neighborhood from M$_{\rm V}$ = 6
$\rm-$ 10. The cluster NGC 6819 is 10 times older than its
relaxation time and therefore has lost some of its low-mass, faint
stars (see Paper II).  

\subsection{Mass Segregation}\label{masssegregation}

    Mass segregation is a consequence of dynamical evolution where
evaporation and redistribution of low mass stars may have occurred 
in the cluster.  
Although this process has been known to occur in open clusters as
early as 1960 (van den Bergh \& Sher), one of the first efforts to
catalogue dynamical effects in many open clusters occurred when
Francic investigated mass functions of eight clusters
\citep{francic}. Francic's study clearly showed that the effects
of mass segregation are more prominent in older clusters. Other
recent studies have also confirmed that the mass functions for
some open clusters have likely changed over time due to dynamical
evolution (for example the Hyades \citep{reid}; NGC 188
\citep{vonHippel2}; Praesepe and NGC 6231 \citep{raboud}; NGC 2420
\citep{lee2}; NGC 2516 \citep{hawley}; M11 \citep{sung}; 
M35 \citep{barrado}; the Pleiades \citep{adams}; NGC
6819 \citep{kalirai2}). Some of these clusters are quite young,
such as the Pleiades, but others such as NGC 2420 and NGC 6819 are
several billions of years old. We should also mention that
\cite{sagar} looked at mass segregation effects for five, distant
open clusters and found that the effects are not correlated with
cluster age. NGC 2099 is an ideal cluster for dynamical studies as
it is neither very old nor very young but rather splits the above
sample in age.  It is also an excellent candidate as it is several
times richer than most of the above clusters.

    A good method to test these evolutionary effects is to compare
the luminosity (or mass) functions in increasing concentric annuli
from the center of the cluster. We split the cluster into four
annuli, each 3$\farcm$5 in radial extent, with geometry summarized
in Table 5. Figure \ref{masssegfig} displays luminosity functions
in each annulus, scaled to the number of stars at M$_{\rm V}$ = 5.
Although not as prominent as in NGC 6819, there is some evidence
for mass segregation of the faintest stars in NGC 2099
(cluster age $\sim$1.7$\times$dynamical age). The outer annuli of
the cluster show a greater relative concentration of lower mass
stars (0.40 M$_\odot$ $\rm-$ 0.60 M$_\odot$).

\section{Mass Functions}\label{massfunc}

    We have used the slope of the M$_{\rm V}-$mass
relation from our theoretical isochrone (520 Myrs) to convert the
observed luminosity function into a mass function in an identical
manner to that described in Paper II for NGC 6819. The mass function 
represents the number of stars / unit mass in the cluster. Typically, 
the mass function is assumed to be a power law so that

%-----------------------------
\begin{equation}
\Psi(m) \propto m^{-(1+x)}, \label{salpeter}
\end{equation}

%-----------------------------
\noindent where $x$ takes on a value of 1.35 in the work of
\cite{salpeter}. For the first cluster in our survey, NGC 6819, we
found that the best fit power law slope was $x$ = $\rm-$0.15, i.e.
the mass function is very flat.  This was expected as the cluster
was 10 times older than its relaxation time.  We can also
summarize the results of \cite{francic} who demonstrated that the
mass functions for some old Galactic clusters (NGC 6633, NGC 752,
and M67) are dominated by higher mass stars. This analysis also
showed that the slope of the mass function for younger open
clusters was $x \sim$ 1. The inverted mass function for the older
clusters is thought to be due to dynamical processes in the
cluster that work to evaporate low mass stars and retain the
higher mass stars. We have shown that it is likely that these
processes may have already occurred in NGC 2099 (see \S\S
\ref{lumfunc} and \ref{masssegregation}). NGC 2099 is only about
1.7 times older than its dynamical relaxation time, and therefore
should exhibit a much steeper global mass function than what was
observed in NGC 6819 if all clusters form initially with a mass
function slope near that of a Salpeter value. Figure
\ref{massfuncfig} shows the mass function for the cluster (solid)
and a Salpeter slope (dashed). This mass function extends from
$\sim$0.3 M$_\odot$ (the limit of our photometry) to $\sim$2.6
M$_\odot$ (the main-sequence turn-off) and has a best fit slope of
$x$ = 0.60 between $\sim$0.5 M$_\odot$ and $\sim$1.8 M$_\odot$,
somewhat flatter than the Salpeter value. The high mass end of the
mass function (between $\sim$1.8 M$_\odot$ and $\sim$2.6
M$_\odot$) is much steeper with a best fit slope of $x$ = 2.5.  The
two lowest mass bins ($\sim$0.3 M$_\odot$ to $\sim$0.5 M$_\odot$)
of the mass function show a very flat distribution.  For such low
masses, color transformation equations are not available for the
non-grey boundary conditions used in our models, however, we have
extrapolated a mass-luminosity relation (as discussed in \S 
\ref{models}).  Integrating the global mass function, and
accounting for the evolved stars, provides a total cluster mass of
$\sim$2300 M$_\odot$.

\section{Star Counts Down to the H-burning Limit}\label{faintmsstars}

    Both the estimated total cluster population ($\sim$2600 stars)
from \S \ref{lumfuncanalysis} and the total cluster mass
($\sim$2300 M$_\odot$) from \S \ref{massfunc} are lower limits due
to our photometric limit (V $\sim$ 23.5 on the main-sequence).  To
obtain a more realistic measure of the total cluster population
(which is required in the estimates of the dynamical time and
tidal radius), we extrapolate our observed luminosity function to
the hydrogen burning limit.  The counts between our faintest bin
($\sim$0.30 M$_\odot$) and the hydrogen burning limit (0.08
M$_\odot$) are obtained by normalizing the Pleiades luminosity
function \citep{lee} to the NGC 2099 function (see Figure
\ref{lumfuncfigcomp}).  The results provide an additional 1350
stars with a total mass of 210 M$_\odot$.  This raises the NGC
2099 cluster population to just under 4000 stars and a total
cluster mass of just over 2500 M$_\odot$.  We also note that the
total mass of the cluster may be larger due to the presence of
binary stars.  The total cluster population may also be higher due
to the missed stars between our cluster extent and the tidal
radius.

\section{White Dwarfs in Open Clusters}\label{wds}

    As mentioned in \S \ref{intro}, observations of white dwarfs
in Galactic open clusters have been limited for several reasons.
The only major large scale observational program to attack issues
such as the initial-final mass relationship and upper mass limit
to white dwarf production in open clusters has been the almost two
decade-long study of Dieter Reimers and Detlev Koester (see
Reimers \& Koester 1988b for an early summary).  The study of these
two relations is very important in order to better understand both
post main-sequence evolution and the chemical evolution of the
disk of our Galaxy. For example, the initial-final mass
relationship depends on the history of stellar mass loss in post
main-sequence evolutionary stages.  The techniques to be used to
establish constraints on these two relations are a lot simpler
than other semi-empirical methods due to large observational and
theoretical uncertainties.  For example, neutron star birth rates
versus white dwarf birth rates can be used to determine the upper
mass limits for these stars, as can the statistics of supernova
explosions.

We will use our rich CMD and white dwarf cooling models to
establish a statistical location for white dwarfs in our CMD.  The
cooling models will yield a preliminary value for the final masses
of the stars and, when combined with the luminosity of the star,
will provide a white dwarf cooling age.  We can then use the
cluster age to determine the main-sequence lifetime for each of
the progenitor stars to the white dwarfs. Finally, main-sequence 
stellar evolutionary models will be used to provide 
initial masses for these stars. The largest uncertainty in the
above analysis is determining accurately the final masses of the
white dwarfs. First, we do not know exactly which objects are
bona-fide cluster members, and secondly, the objects do not
closely follow any particular mass cooling sequence.  The current
project will therefore identify possible candidates that will then
be spectroscopically observed using multi-object spectrographs on
8-meter class telescopes to isolate cluster members and measure
surface gravities and effective temperatures (which gives a more
accurate value of the final mass).  This is a similar approach to
that of Reimers and Koester who used spectroscopy to confirm and
identify bright white dwarfs (individually) in sparsely populated
open clusters that had been discovered in previous published
imaging projects.  Only a few objects were found in each study and
the results for all clusters were combined to establish
constraints on the relationships outlined above. Among the
clusters studied were NGC 2516
\citep{reimerskoester8,reimerskoester5}, NGC 6633
\citep{reimerskoester1}, NGC 3532 \citep{reimerskoester2}, NGC
2168 \citep{reimerskoester3}, IC 2391 and NGC 2451
\citep{reimerskoester4}, and NGC 2287 and NGC 2422
\citep{reimerskoester6}.

    The clusters studied by Reimers and Koester do not generally contain
enough white dwarfs to accurately establish a cooling sequence and
therefore, an independent age measurement. However, white dwarf
cooling ages have been established for older open star clusters
such as M67 \citep{richer2} and NGC 2420 \citep{vonHippel1}. The
difficulties in these studies are that the photometry does not
extend much fainter than the end of the cooling sequence and that
the termination of the white dwarf cooling sequence is buried in
field stars or unresolved galaxies making it very difficult to
actually isolate it on the CMD. The statistics in these clusters
do show that the end of the cooling sequence has been detected,
but these are affected by large errors.

    Clusters such as NGC 2099 and NGC 6819 (Paper II) possess
advantages over both the young clusters studied by Reimers and
Koester and these older clusters.  NGC 2099 is several times
richer than any of the younger clusters and the photometry is both
more accurate and deeper. Therefore, spectroscopic observations of
the objects in this cluster will not only increase the sample size
by many factors, but also establish constraints from fainter,
older white dwarfs.  The cluster is also approximately three times
richer than the richest of the older clusters for which a white
dwarf cooling age has been established. The white dwarf population
is also well separated from the bulk of the field stars thereby
making the statistics simpler. Additionally, our photometry
extends almost 1.5 magnitudes fainter than the termination of the
white dwarf cooling sequence and the cluster CMD clearly shows
that we have detected the end of the cooling sequence.  A
disadvantage of studying white dwarfs in NGC 2099 rather than M67,
NGC 2420 and NGC 6819 is that the cluster is not yet old enough
for a long trail of white dwarfs to have formed so it is difficult
to fit a cooling model over a wide magnitude range of white
dwarfs.

\subsection{Removing Non-cluster Objects}
\label{wdcontamination}

    The background field CMD (Figure \ref{cmds2099}) shows that
there is a significant population of field stars and/or 
faint background galaxies in the region where cluster white dwarfs
are expected (see 22 $\leq$ V $\leq$ 25, 0 $\leq$ B$\rm-$V $\leq$
0.6). However, this population is also clearly more dense in the
cluster field. Although we can not delineate those objects that
are cluster members from the CMD alone, we can use the background
field to statistically determine the number of expected cluster
members and their most likely location in the CMD.  We count all
objects in the background field within each 0.50 magnitude interval
in V and remove 1.37 times the number of field objects in the same
magnitude bin. This approach is only invoked for possible white
dwarfs in the faint, blue end of the CMD.  It is important to note
that we may in fact remove all white dwarf candidates in this
process if there were no cluster white dwarfs. This statistical
technique no longer allows us to consider individual objects as
potential white dwarfs: multi-object spectroscopy of all potential
candidates, before any cuts, with instruments such as GMOS on
Gemini North or LRIS on Keck, will provide the definitive answer
as to which objects really are white dwarfs. If we assume that the
clustering properties of background galaxies is almost uniform
across the 42$'$ $\times$ 28$'$ field, then this statistical
subtraction should also eliminate the correct number of galaxies
from the cluster field.

    It is difficult to estimate which stellarity cut is an optimal
separation of galaxies from stars, so we must consider the
expected number of galaxies in our cluster field (R $\leq$
13$\farcm$9).  This is done by considering galaxy counts
\citep{woods} at high latitude and correcting for extinction in
our field.  Based on these statistics and prior to any stellarity
cut, there are far more objects in the brightest of our faint
magnitude bins than the number of expected galaxies. For example,
we expect less than 15\% of all objects to be galaxies for 21
$\leq$ V $\leq$ 22. However, for 22 $\leq$ V $\leq$ 23, we expect
37\% of our objects to be galaxies, and for 23 $\leq$ V $\leq$ 24
fully 98\% of our objects are expected to be galaxies. Remarkably,
invoking a 0.50 stellarity cut removes approximately the correct
number of galaxies from two of these three bins. For 21 $\leq$ V
$\leq$ 22, we expect 336 galaxies in our cluster field and we
remove 430 objects with a 0.50 stellarity cut. For the 22 $\leq$ V
$\leq$ 23 bin, the number of objects removed at a 0.50 stellarity
cut is 1015 and the number of galaxies is 1016.  For 23 $\leq$ V
$\leq$ 24, the number of objects removed at a 0.50 stellarity cut
is 1774 and the number of expected galaxies is 2552.  A 0.75
stellarity cut in this last bin is almost perfect in eliminating
the correct number of objects. It eliminates 2556 of the expected
2552 galaxies.  Perhaps after the study of many clusters in the
survey we can establish with certainty the correct stellarity cut
for separating galaxies from stars, at a given signal-to-noise
ratio.  For this work, we will use a 0.50 stellarity cut which is
the same result found for the photometry of NGC 6819 in Paper II.

After statistical subtraction as described above, most of the 
remaining objects in the faint-blue end of the 
CMD should be white dwarfs.  There may also be some unresolved
galaxies or field blue objects.  We can estimate the number of
expected field white dwarfs and compare this value to the number
of faint-blue objects in our field.  We use the hot white dwarf
luminosity function (Figure 2 in Leggett, Ruiz \& Bergeron 1998, 
taken from Liebert, Dahn \& Monet 1988) to count the number of 
expected white dwarfs per 
pc$^{3}$ above our limiting magnitude. First, a white dwarf
cooling model for 0.70 M$_\odot$ \citep{wood} is used to convert
our absolute magnitude bins to bolometric magnitude bins. This
then provides a value for the bolometric luminosity of the white
dwarfs at each bin center. The number of expected white dwarfs,
above that luminosity in that bin, simply follows from the Figure.
Adding up the numbers in the bins from M$_{\rm V}$ = 10.5 to 13
and multiplying by the volume created by the cone that represents
our cluster field, yields a population of 63 expected field white
dwarfs.  The number of objects in our background field, after
correcting for incompleteness in our data and removing background
galaxies, is 35, which when scaled up to match the size of the
cluster field gives 48.  The difference between the expected and
the observational values can be most likely attributed to errors
in the galaxy counts and therefore in the stellarity cut used and
also to white dwarfs that are in binaries and therefore missed in
our study (see next section).

\subsection{White Dwarfs in Binaries}
\label{binaries}

    Evidence supporting a population of white dwarfs in
binaries in NGC 2099 is provided from a very crude estimate of the
expected number of white dwarfs in the cluster based on an
extrapolation of the observed mass function at the high mass end.
As mentioned in \S \ref{massfunc}, the high mass end slope of the
mass function is $x$ = 2.5.  Using the number counts in the m =
2.22 M$_\odot$ mass bin to solve for the proportionality constant
(A = 3684), and integrating equation \ref{salpeter} from the
turn-off (M $\sim$ 2.6 M$_\odot$) to the upper limit to white
dwarf production (M $\sim$ 7 M$_\odot$) provides a value for 
the expected number of white dwarfs in the cluster to be 
113.  This value is $\sim$2 times greater than the number of
detected white dwarfs (see next section).  A significant number of
these missing white dwarfs are most likely in main-sequence
binaries, however, a few additional arguments can also help
explain the discrepancy.

    First, the cluster will lose stars through tidal stripping and
evaporation.  The white dwarfs may be more prone to these effects
because of the planetary nebulae stage.  Any asymmetric
mass-ejection would induce a recoil velocity on the white dwarf
that could kick it out of the cluster.  Another effect that could
possibly lead to fewer detected white dwarfs is relaxation (see \S\S 
\ref{lumfunc} and \ref{masssegregation}).  Since the white dwarfs 
have a small mass compared to their progenitors, relaxation could 
increase their velocity dispersion and therefore make them less 
frequent in the cluster center (compared to the progenitors) and 
more susceptible to evaporation.  This effect could not however 
be very dramatic as the white dwarf cooling ages for most of the 
candidates are comparable to the relaxation age for the cluster.

\subsection{White Dwarf Luminosity Function and Cooling Age}
\label{wdcoolingage}

    For those objects that have survived the criteria outlined in
\S \ref{wdcontamination}, we construct a white dwarf luminosity
function by binning the objects in 0.50 magnitude intervals and
subtracting the background field numbers from the cluster field
numbers after accounting for the 1.37 aerial difference.  Figure
\ref{wdlumfunc} shows this result, after accounting for
incompleteness corrections as described in \S \ref{incompcorr} and
summarized in Table 3. There is a clear increase in the number of
stars as a function of magnitude and then a sharp turn-off at V =
23.5 (M$_{\rm V}$ = 11.95 $\pm$ 0.30) (see Table 6).  The
uncertainty in this value is discussed in detail in the next
section.  We interpret this turn-off to represent the end of the
white dwarf cooling sequence in NGC 2099.  The total number of
white dwarfs in NGC 2099, found from adding up the bins in the
luminosity function, is 50. The bright magnitude slope of the
luminosity function is in excellent agreement with the expected
slope of the white dwarf luminosity function from theory
\citep{liebert}. We can use the magnitude of the termination point
of the white dwarf cooling sequence to establish a lower limit to
the age of NGC 2099 based on white dwarf cooling models
\citep{wood}.  Figure \ref{3plots} shows a 0.70 M$_\odot$ cooling
model superimposed on the cluster CMD.  The age from the limiting
magnitude in this cooling sequence is 516 $\pm \ ^{154}_{176}$
Myrs.  This age is a lower limit because it represents the time
taken for the most massive progenitor stars in the cluster to cool
to the limiting white dwarf magnitude and does not include the
time that these stars spent on the main-sequence. The latter, for
7 M$_\odot$ stars, is 50 Myrs, therefore raising the white dwarf
cooling age of the cluster to 566 $\pm \ ^{154}_{176}$ Myrs which
is in excellent agreement with the core-overshooting turn-off age
(520 Myrs). Finally, we note that if we assume a slightly higher
mass for the white dwarfs by 0.10 M$_\odot$, the age decreases
slightly (530 Myrs) and is in better agreement with the turn-off
age. Surprisingly, the cooling ages determined here are
insensitive to the stellarity cut (0.50 or 0.75).  The peak in the
white dwarf luminosity function occurs at M$_{\rm V}$ = 11.95
independent of a 0.50 or 0.75 stellarity cut. The number of
objects however, will clearly be less in the 0.75 case.  This is
in part due to the resolution of the bins used for the white dwarf
luminosity function.

\subsection{Evaluation of Errors}
\label{uncertainty}

    In order to compare the white dwarf cooling age with
the main-sequence turn-off age, we must provide a careful account
of all errors.  As mentioned in \S \ref{parameters}, the total
error contribution to the true distance modulus includes four
terms. Additionally, the error in the limiting white dwarf cooling
magnitude must also include the bin uncertainty ($\pm$0.25
magnitudes) from Figure \ref{wdlumfunc}. We can address each of
the uncertainties individually and then combine them
correctly to account for correlations between the cross terms: \\

1.) There is an error in the vertical shift caused by the error in
the horizontal shift due to $\Delta_{E(B-V)}$.  This error is
proportional to $\Delta_{E(B-V)}$ and to the slope of the
main-sequence at the fitting point ($S$ = $\Delta V$/$\Delta (B-V)
\sim$ 5.7).  Therefore, $\Delta_{A}$ = $S\Delta_{E(B-V)}$ = 0.17.

2.) There also exists a judgement uncertainty in the vertical shift needed
to superimpose the Hyades main-sequence on the NGC 2099 main-sequence
at a fixed value of E(B$\rm-$V).  Under some circumstances, this shift
is very hard to determine due to large photometric spread in the
main-sequence and the smearing effect of binaries.  Given the tightness
of the two main-sequences, we estimate this error to be quite small,
$\Delta_{B}$ = 0.07.

3.) The third error we consider is the error in the extinction
($\Delta_{A_{V}}$) caused by the uncertainties in both the
reddening ($\Delta_{E(B-V)}$) and $\Delta_{R_{\rm V}}$.  This term
is evaluated as

%-----------------------------
\begin{equation}
\Delta_{C} = \Delta_{A_{V}} = A_{V}\sqrt{(\frac{\Delta_{R_{V}}}{R_{V}})^2 +
(\frac{\Delta_{E(B-V)}}{E(B-V)})^2}, \label{extinctionerror}
\end{equation}

%-----------------------------

\noindent where we take $\Delta_{R_{\rm V}}$ = 0.20.  The above equation
gives $\Delta_{C}$ = 0.10.

4.) The final term in our distance modulus error budget accounts for the
small color shift caused by the metallicity uncertainty of the cluster.
As described in \S \ref{metallicity}, the metallicity is believed to be
Z = 0.02. If we assume an uncertainty $\Delta_{Z} \sim$ 0.005 (this allows
the cluster metallicity to be equal to that of the Hyades), then the
corresponding shift in color is $\Delta_{B\rm-V}$ = $\Delta_{D}$ = 0.03.

We combine the above errors by writing the total error as a sum of these
terms and then taking the mean and root as shown here:

%-----------------------------
\begin{eqnarray}
\Delta_{(m-M)_{\rm o}} & = & \langle(\Delta_{A} + \Delta_{B} + \Delta_{C} +
\Delta_{D})^{2}\rangle^{1/2} \nonumber \\ & = & \langle\Delta_{A}^{2} + \Delta_{B}^{2} + \Delta_{C}^{2} + \Delta_{D}^{2} + 2\Delta_{A}\Delta_{B} + 2\Delta_{A}\Delta_{C} \nonumber \\ & & \mbox{} + 2\Delta_{B}\Delta_{C} + 2\Delta_{A}\Delta_{D} + 2\Delta_{B}\Delta_{D} + 2\Delta_{C}\Delta_{D}\rangle^{1/2}. \label{errorestimate}
\end{eqnarray}

%-----------------------------

Next, we can eliminate three of the six cross terms,
2$\Delta_{A}\Delta_{B}$, 2$\Delta_{B}\Delta_{C}$ and
2$\Delta_{B}\Delta_{D}$, because they are the products of
un-correlated errors; they are defined at fixed E(B$\rm-$V)
so $\Delta_{E(B-V)}$ does not enter into these terms.  Finally,
we note that one of the cross terms, 2$\Delta_{A}\Delta_{C}$,
is negative because the two errors compensate for each other.
An error in $\Delta_{E(B-V)}$ in one of the terms opposes the
associated error caused by the $\Delta_{E(B-V)}$ in the other
term.  Evaluating equation \ref{errorestimate} with these
modifications gives a total error value of 0.16 on the true
distance modulus.

The error in the limiting white dwarf magnitude can be found by
simply adding this distance modulus error ($\pm$0.16) in quadrature
with the binning error ($\pm$0.25).  The result gives $\pm$0.30.

\subsection{Summary of White Dwarf Studies}
\label{wdsummary}

    This project has allowed us to identify 50 white dwarf
candidates in NGC 2099.  These objects have all passed a
stellarity and a statistical subtraction cut.  Furthermore, the
bright end slope of the white dwarf luminosity function agrees
well with theoretical expectations.  The turn-over in the white
dwarf luminosity function provides an age measurement for the
cluster (566 $\pm \ ^{154}_{176}$ Myrs) which is in excellent agreement
with the main-sequence turn-off age (520 Myrs) for a core-overshooting
model. A non core-overshooting age of $\sim$300$\rm-$350 Myrs for
NGC 2099 can therefore be ruled out with some certainty. The
approach that we have used to estimate these quantities is a
purely statistical method that needs further work.  Fortunately,
new multi-object spectrometers will allow us to simultaneously
measure spectra for a large number of objects. We will eventually
obtain spectra for all objects in the faint, blue end of the CMD
before any cuts have been made. This will then provide a true
measure of the number of white dwarfs in NGC 2099.  Accurate
surface gravities and effective temperatures for the brighter
objects will provide crucial mass information which, when coupled
with theoretical models, provides progenitor mass information.
This will lead to a better understanding of the initial-final mass
relationship for white dwarfs.

%-----------------------------

\section{Conclusion} \label{conclusion}

    We have used deep CFH12K photometry to establish the white
dwarf cooling age of one of the richest open star clusters, NGC
2099.  The white dwarf luminosity function for the cluster shows a
steep turn-over at M$_{\rm V}$ = 11.95 $\pm$ 0.30, which is the
limit to which the most massive progenitor stars have cooled to.
After accounting for the main-sequence lifetime of these stars,
the cooling age is determined to be 566 $\pm \ ^{154}_{176}$ Myrs,
and therefore is in excellent agreement with the main-sequence
turn-off age (520 Myrs).  In order to have consistency between the
two ages, core-overshooting models are preferred. The \it V,
B$\rm-$V \normalfont CMD for NGC 2099 exhibits a spectacularly
rich, long (over 12 magnitudes) and very tightly constrained
main-sequence. Main-sequence fitting of the un-evolved stars in
the cluster with the Hyades cluster indicates a true distance
modulus of (m$\rm-$M)$_{\rm o}$ = 10.90 $\pm$ 0.16 and a reddening
value of E(B$\rm-$V) = 0.21 $\pm$ 0.03. After accounting for small
metallicity differences, the slope changes in the NGC 2099
main-sequence are found to be matched almost perfectly to those in
the Hyades cluster.  A theoretical stellar isochrone with Z =
0.020 and age = 520 Myrs models several features of the CMD very
well, such as slope changes in the upper main-sequence and the red
giant clump. The lower main-sequence and the turn-off are found to
be slightly redder than the isochrone. The radial distibution of
stars in NGC 2099 are in good agreement with a single mass King
model. The global luminosity function is found to be flat from
M$_{\rm V} \sim$ 2 to M$_{\rm V} \sim$ 8 with a subsequent rise
for fainter stars. After accounting for stars between our faintest
magnitude bin and the hydrogen burning limit, the total cluster
population of NGC 2099 is determined to be just under 4000 stars.
Although not as severe as for the older cluster NGC 6819, we find
some evidence for mass segregation within NGC 2099. The global
mass function is shallower than a Salpeter IMF, and when
extrapolated to the hydrogen burning limit provides a lower limit
(excluding binaries and stars out to the tidal radius) to the
total cluster mass of $\sim$2515 M$_\odot$.

%-----------------------------

\acknowledgments

We wish to thank J. Montalban for helpful discussions concerning
the modelling of low mass main-sequence stars and S. Courteau for
helpful discussions concerning error analysis. The author would
also like to thank the Canada-France-Hawaii Telescope Corporation
for a generous three night observing run with their facility and
the CFHT Corporation and Osservatorio Astronomico di Roma for 
accommodations and access to computing facilities during this project.  
The author received financial support during this work through an 
NSERC PGS-A research grant.  The research of HBR and GGF is supported 
by the Natural Sciences and Engineering Council of Canada.

\clearpage

%% No more than seven \figcaption commands are allowed per page,
%% so if you have more than seven captions, insert a \clearpage

%% There must be a \figcaption command for each legend. Key the text of the
%% legend and the optional \label in curly braces. If you wish, you may
%% include the name of the corresponding figure file in square brackets.
%% The label is for identification purposes only. It will not insert the
%% figures themselves into the document.
%% If you want to include your art in the paper, use \plotone.
%% Refer to the on-line documentation for details.

\clearpage

\figcaption[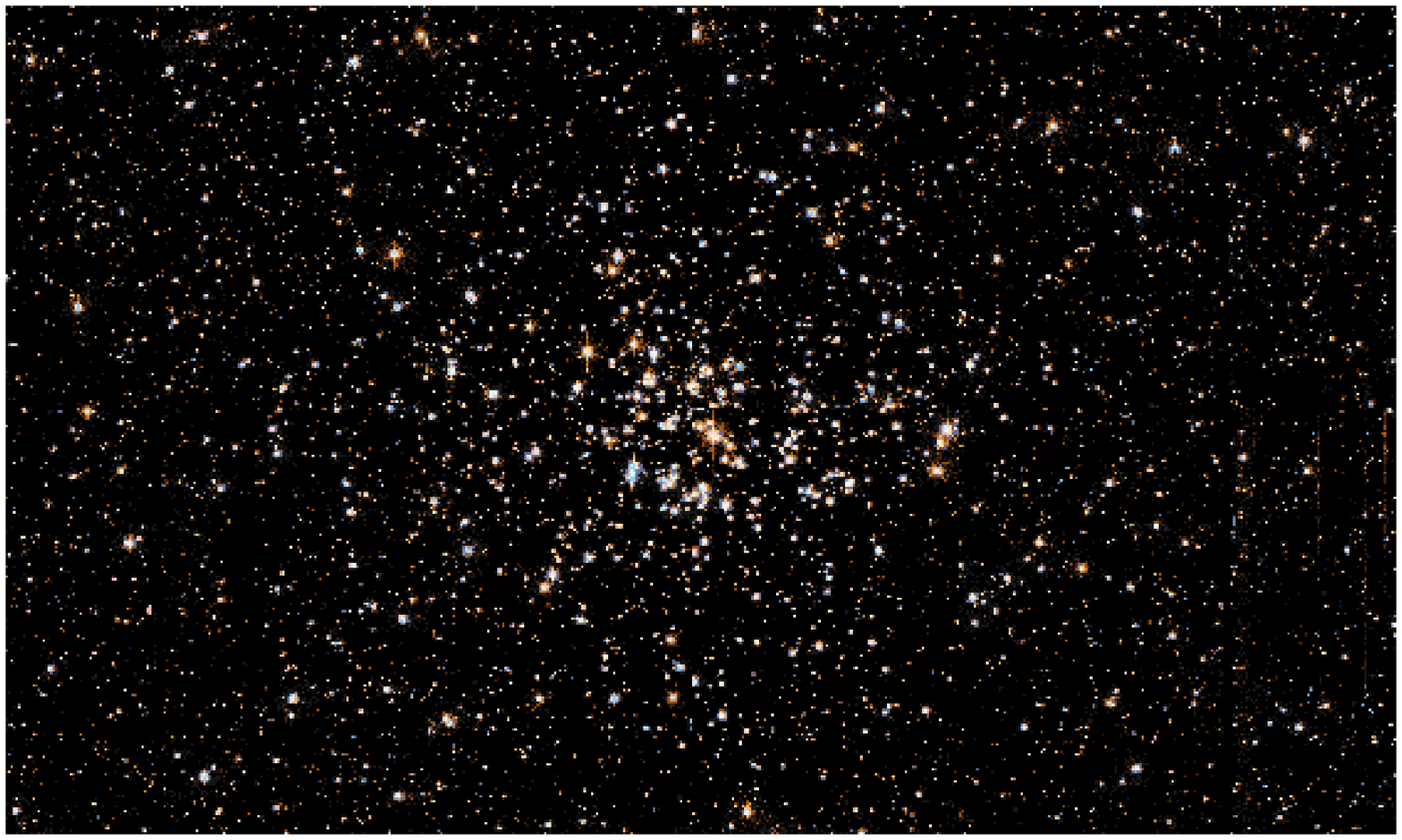]{Color image created from the
individual V and B 300-second frames.  The image size is 42$'$
$\times$ 28$'$. \label{newmosaic}}

\epsscale{0.75}

\plotone{Kalirai.fig1.eps}

\clearpage

\figcaption[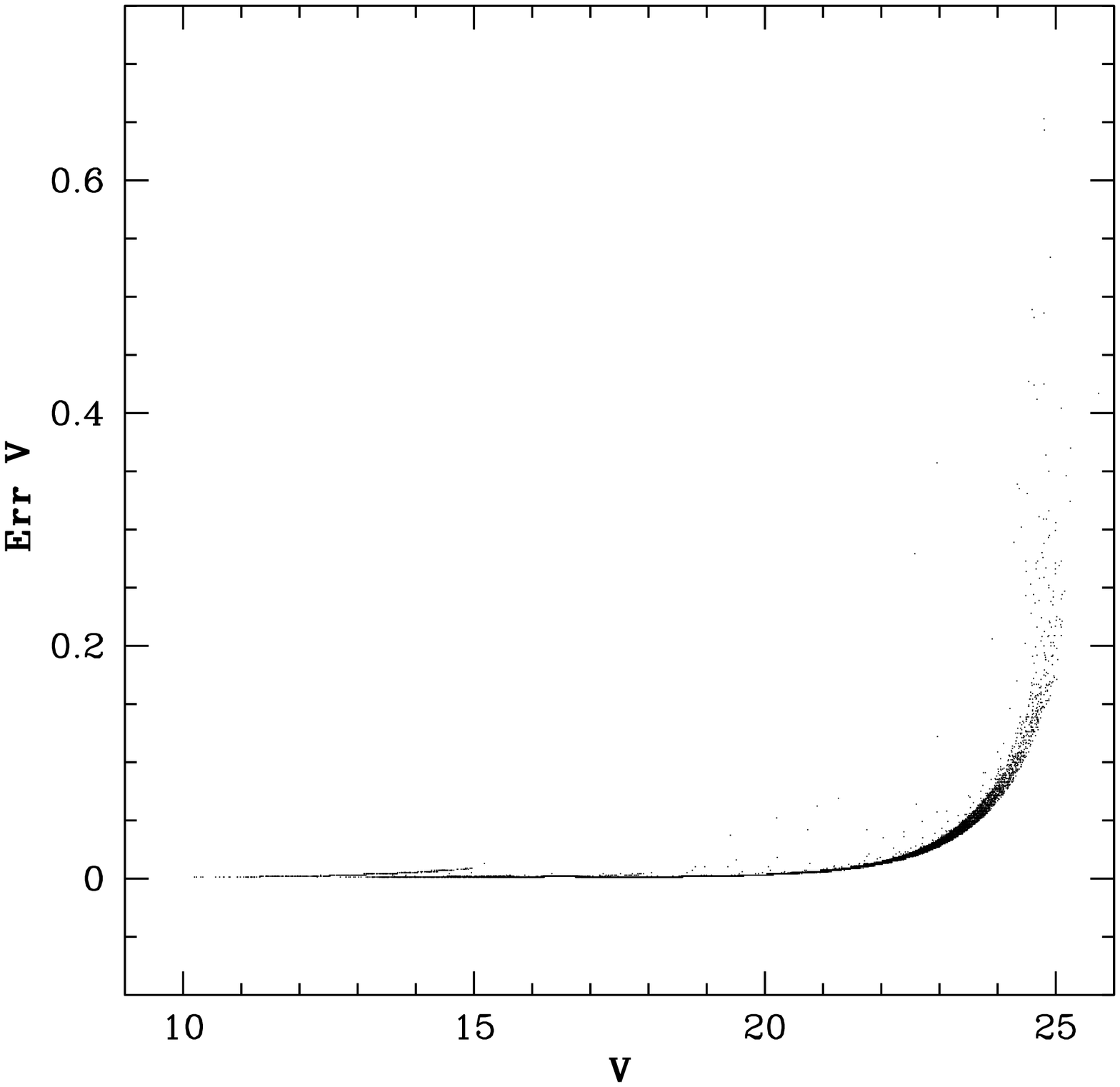]{Statistical errors in the PSFex
photometry.  These are negligible up to V = 22, at which point
they start rising rapidly.  The mean error at V = 23 is 0.032 and
at V = 24 is 0.075. Very few outliers are present in the
photometry.  Two small `glitches' at 13 $\leq$ V $\leq$ 15 and 
V $\sim$ 17.5 indicate the magnitudes at which the different exposure 
time data sets were merged.  \label{errfig}}

\plotone{Kalirai.fig2.eps}

\clearpage

\figcaption[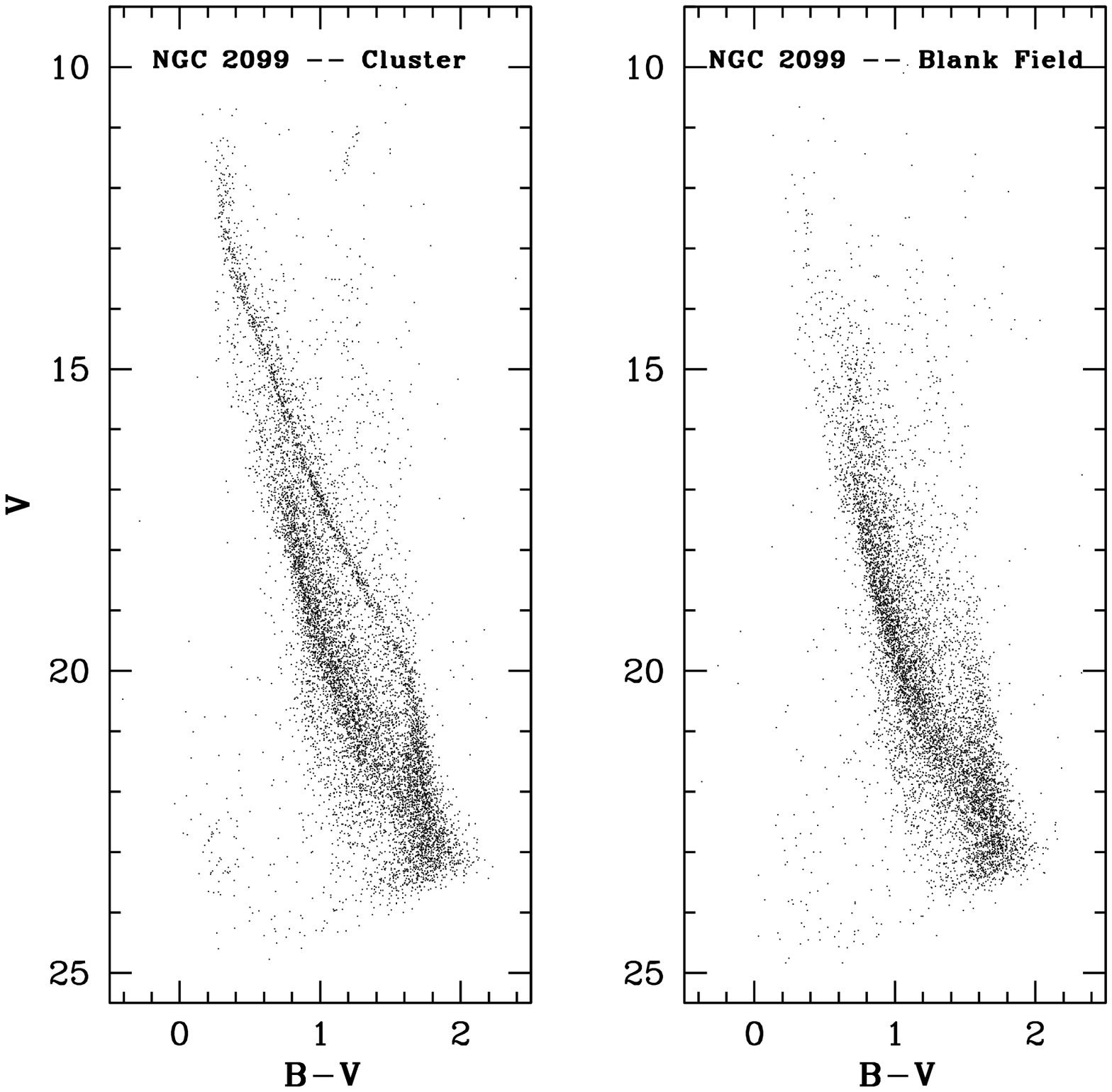]{Rich, tight and long main-sequence
of NGC 2099 is clearly separated from the background/foreground
distribution. The cluster field has been scaled down in area by a
factor of 1.37 (as described in text) so that the relative
populations on the two diagrams can be compared (see Figure
\ref{comparisonfig} for all stars).  A 0.50 stellarity cut has
been applied to both diagrams. \label{cmds2099}}

\plotone{Kalirai.fig3.eps}

\clearpage

\figcaption[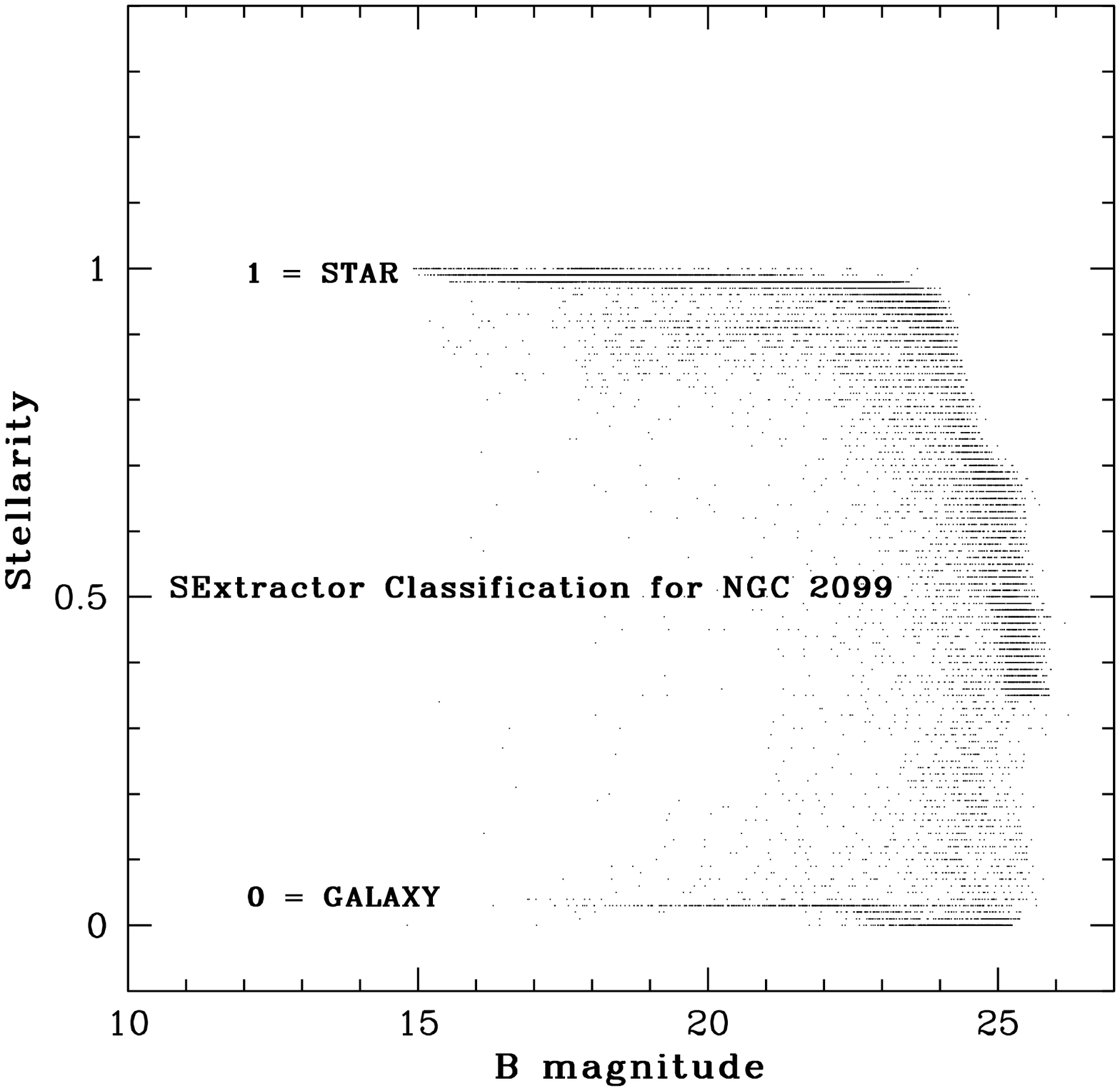]{SExtractor star/galaxy classification 
shown for all objects in the cluster field.  A number of stars
(stellarity = 1) and galaxies (stellarity = 0) are clearly
evident.  The classification of the remaining objects is discussed
in \S\S \ref{stellarity} and \ref{wdcontamination}. \label{stellfig}}

\plotone{Kalirai.fig4.eps}

\clearpage

\figcaption[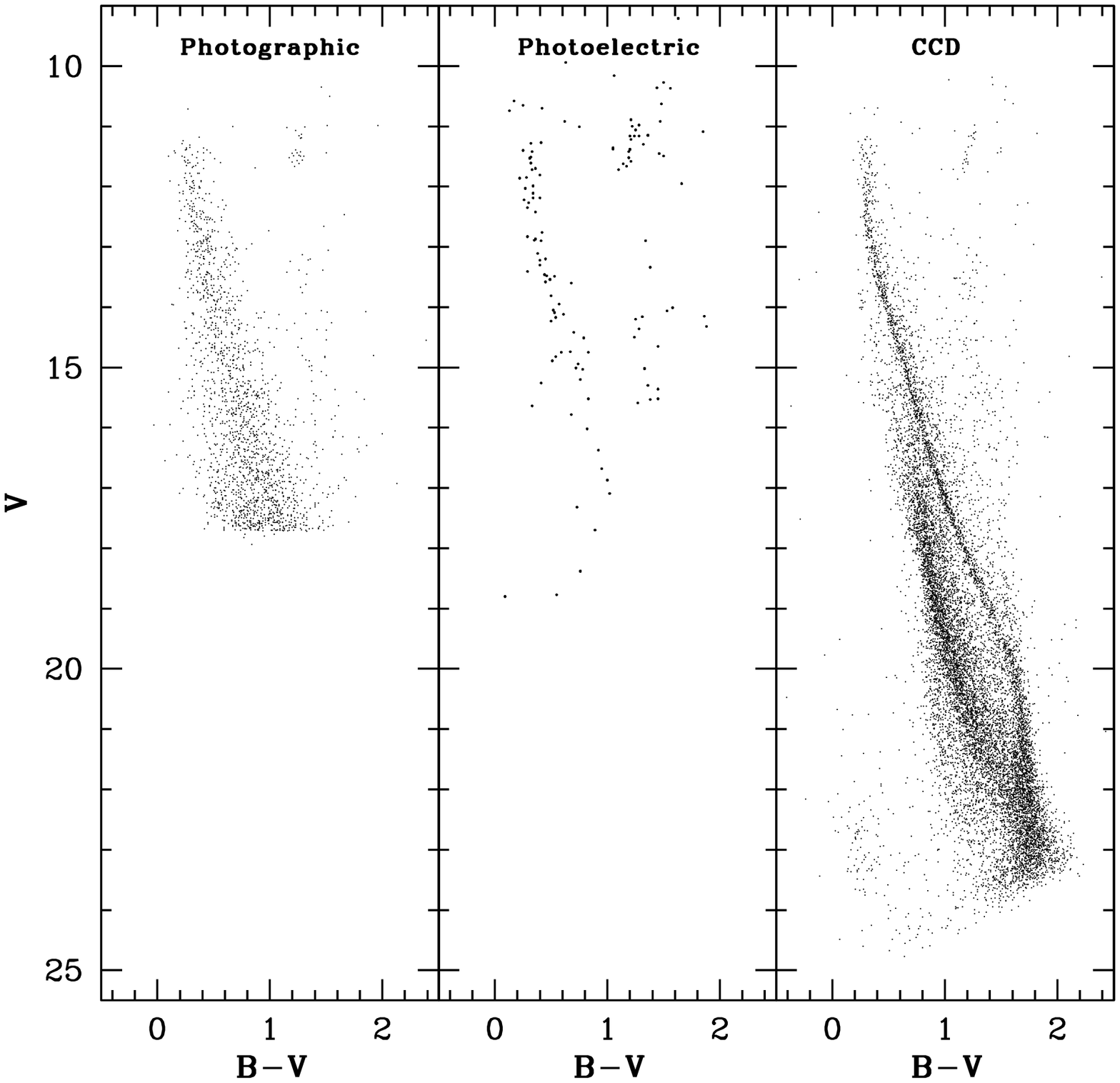]{The present CCD photometry is
compared to the NGC 2099 studies of West (photographic) and
Mermilliod (photoelectric).  Unlike Figure \ref{cmds2099}, the
right hand panel here contains all cluster stars out to R =
13$\farcm$9. \label{comparisonfig}}

\plotone{Kalirai.fig5.eps}

\clearpage

\figcaption[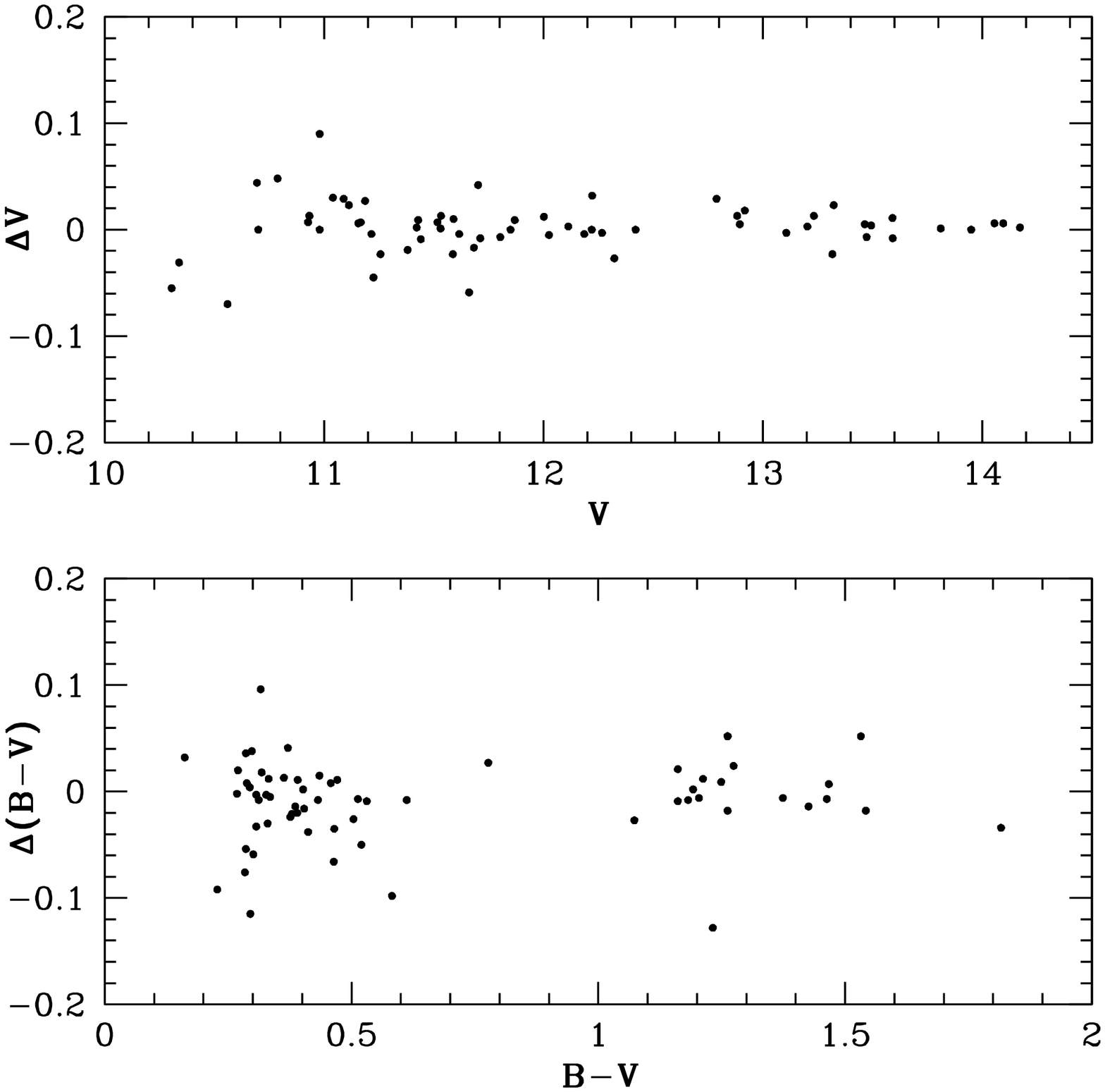]{Photometric magnitudes for bright 
individual stars in the present CCD photometry are compared to 
the values obtained in the Mermilliod et al. (1996) photoelectric 
study.  The residuals (which show CCD $\rm-$ photoelectric vs CCD) 
are found to be very small and show no obvious biases or trends.  
Surprisingly, the residuals are higher for brighter sources which may 
be due to the extremely short exposures required to measure these 
stars in the CCD case. \label{matchfig}}

\plotone{Kalirai.fig6.eps}

\clearpage

\figcaption[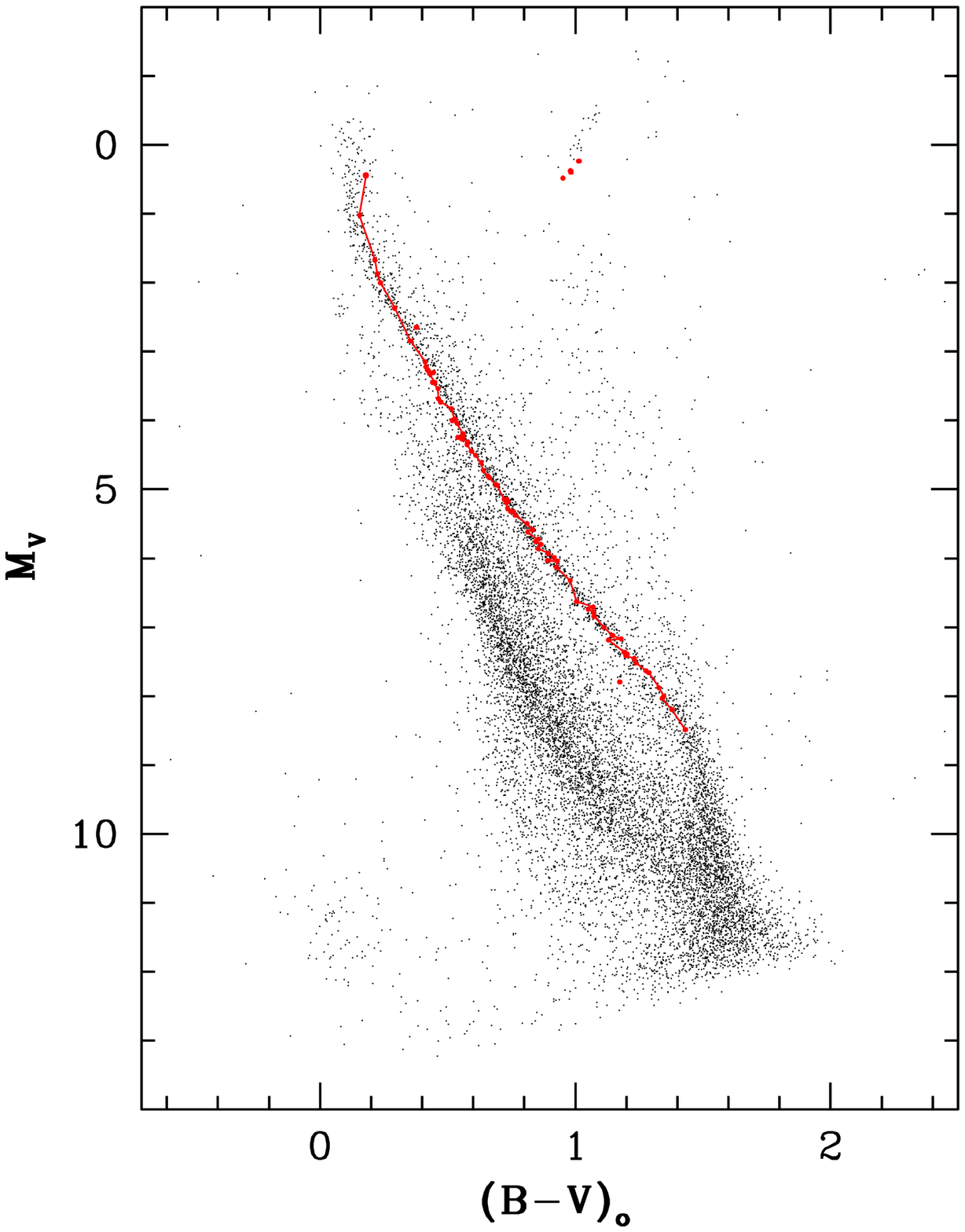]{NGC 2099 main-sequence is brought
to the Hyades (red points) plane by adjusting the metallicity,
reddening and distance as described in \S\S \ref{metallicity} and
\ref{parameters}.  The resulting detailed shape of the NGC 2099
main-sequence is found to be in excellent agreement with the
Hyades cluster, thus simplifying the main-sequence fitting.  The
turn-off and clump stars in the Hyades clearly indicate a slightly
older cluster, as confirmed by the isochrones ($\Delta$age $\sim$
100 Myrs). \label{hyadesfig}}

\plotone{Kalirai.fig7.eps}

\clearpage

\figcaption[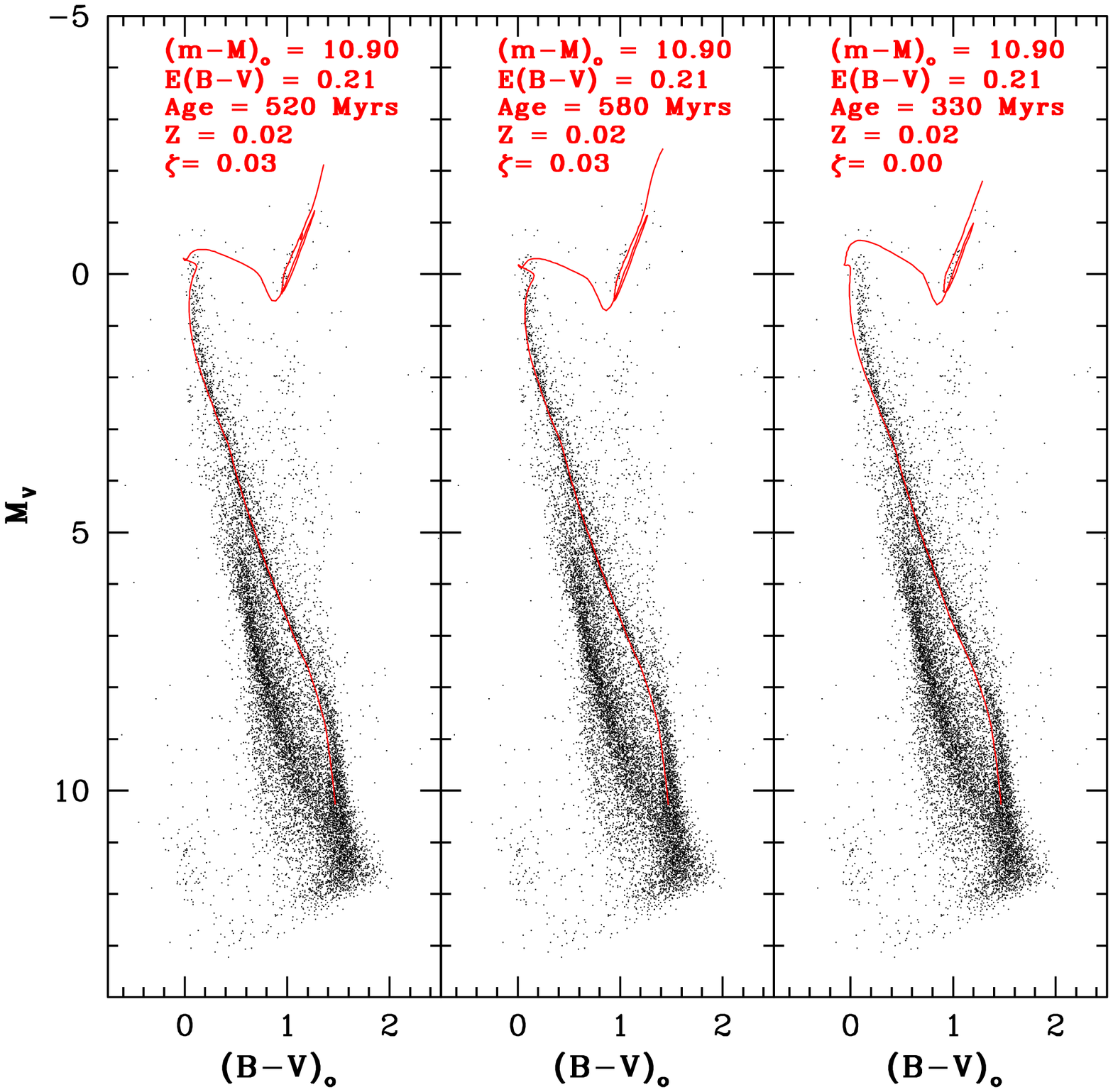]{The left panel shows the best fit
isochrone, based on the luminosity of the clump stars, to the
observed NGC 2099 data.  A better fit to the turn-off of the
main-sequence is shown in the middle panel, however, a close
examination of the clump stars shows a $\sim$0.20 magnitude 
discrepency when compared to this model.  A non core-overshooting 
model (right panel) provides a poor fit to the data. See \S
\ref{obstheory} for a discussion of these results. \label{3isos}}

\plotone{Kalirai.fig8.eps}

\clearpage

\figcaption[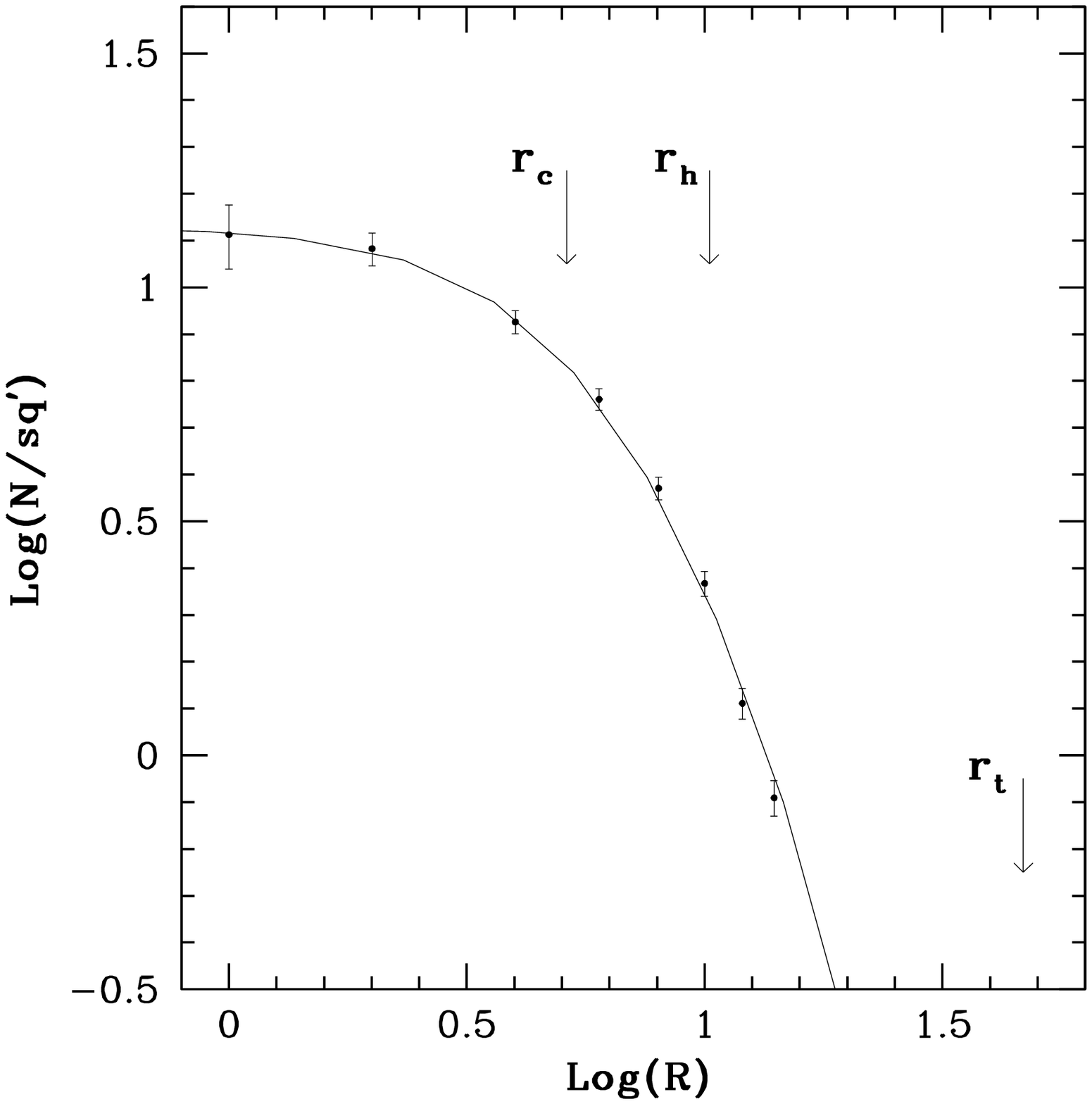]{Single mass King model found to be
in good agreement with the NGC 2099 radial density distribution.
Only stars with V $\leq$ 22 have been counted to avoid
incompleteness effects. The arrows mark the core, half-mass and 
tidal radii of NGC 2099. \label{kingfig}}

\plotone{Kalirai.fig9.eps}

\clearpage

\figcaption[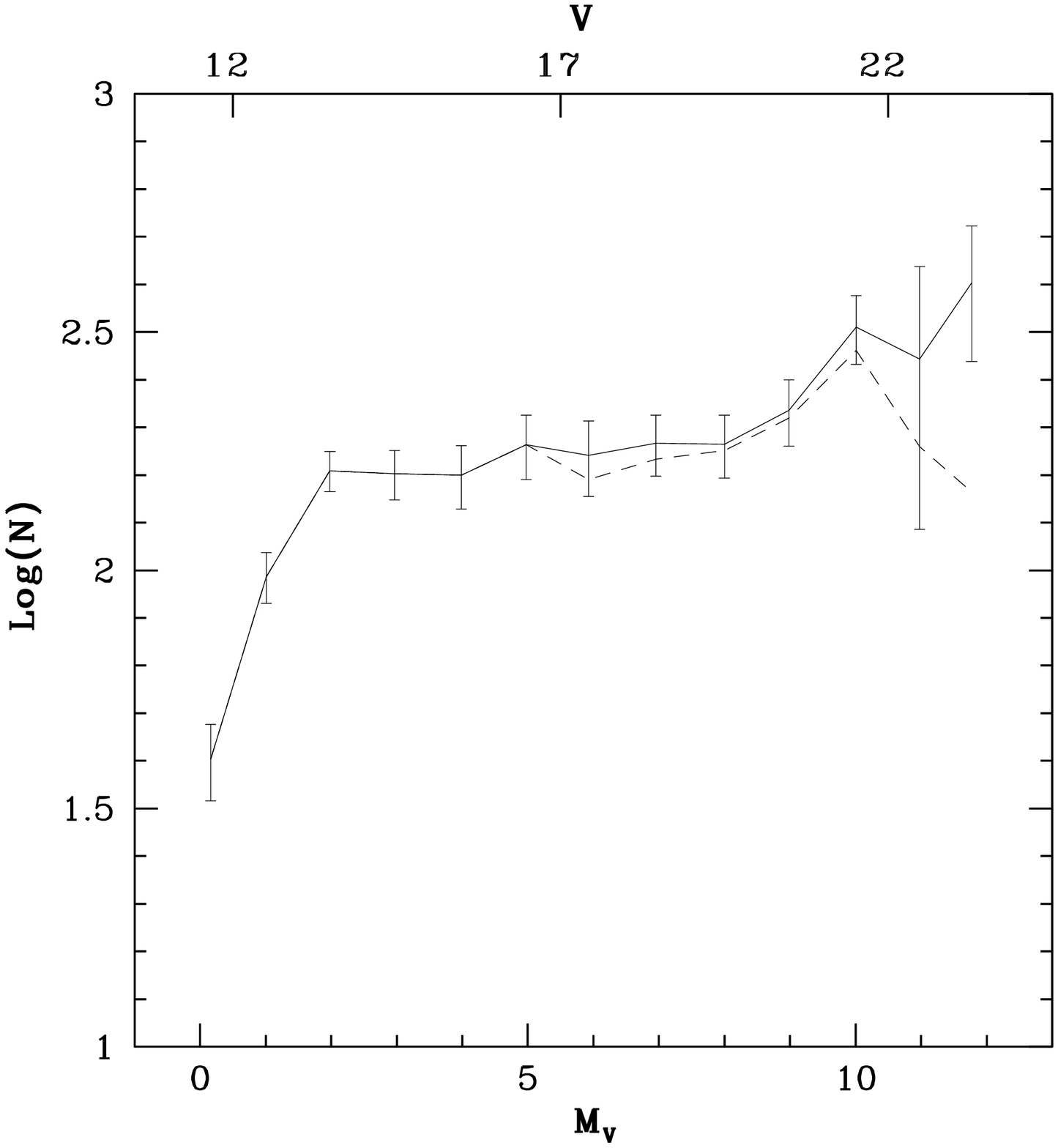]{Global (R $\leq$ 13$\farcm$9)
luminosity function is shown before (dashed) and after (solid)
incompleteness corrections.  The rising portion of the luminosity
function beyond V $\sim$ 19.5 (M$_{\rm V} \sim$ 8) corresponds to a
slope change in the lower main-sequence of the cluster CMD, and is
caused by an inflection in the mass-luminosity relationship.  The
error bars reflect a combination of Poisson errors and
incompleteness errors as discussed in \S \ref{incompcorr}.
\label{lumfuncfig}}

\plotone{Kalirai.fig10.eps}

\clearpage

\figcaption[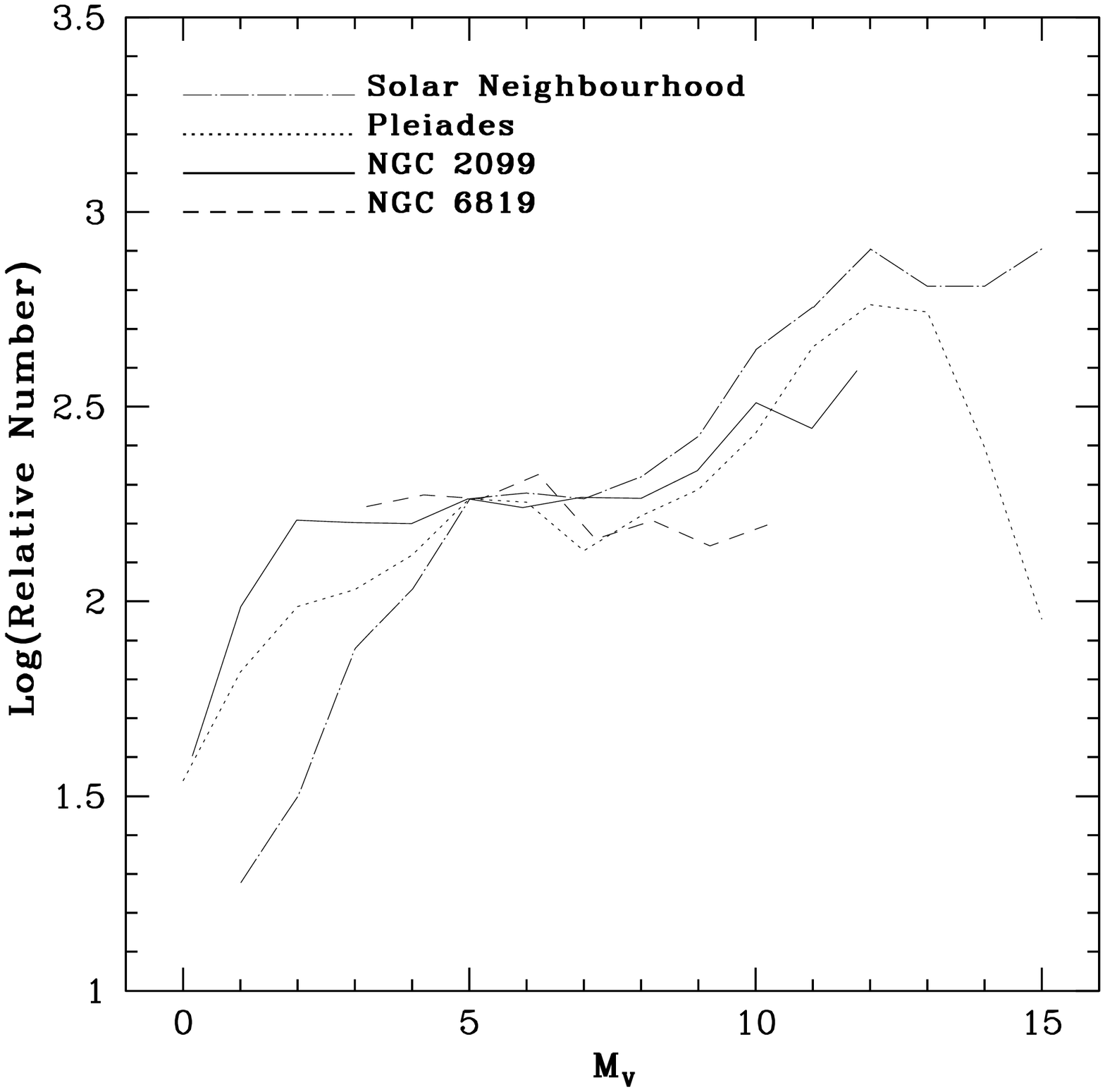]{NGC 2099 luminosity function
is compared with the Solar Neighborhood distribution of stars, the
Pleiades luminosity function and the NGC 6819 luminosity
function. All luminosity functions have been normalized to the
number of stars in NGC 2099 at M$_{\rm V}$ = 5 to avoid
evolutionary effects.  NGC 6819 exhibits an inverted slope with
respect to the other functions due to significant dynamical
effects. \label{lumfuncfigcomp}}

\plotone{Kalirai.fig11.eps}

\clearpage

\figcaption[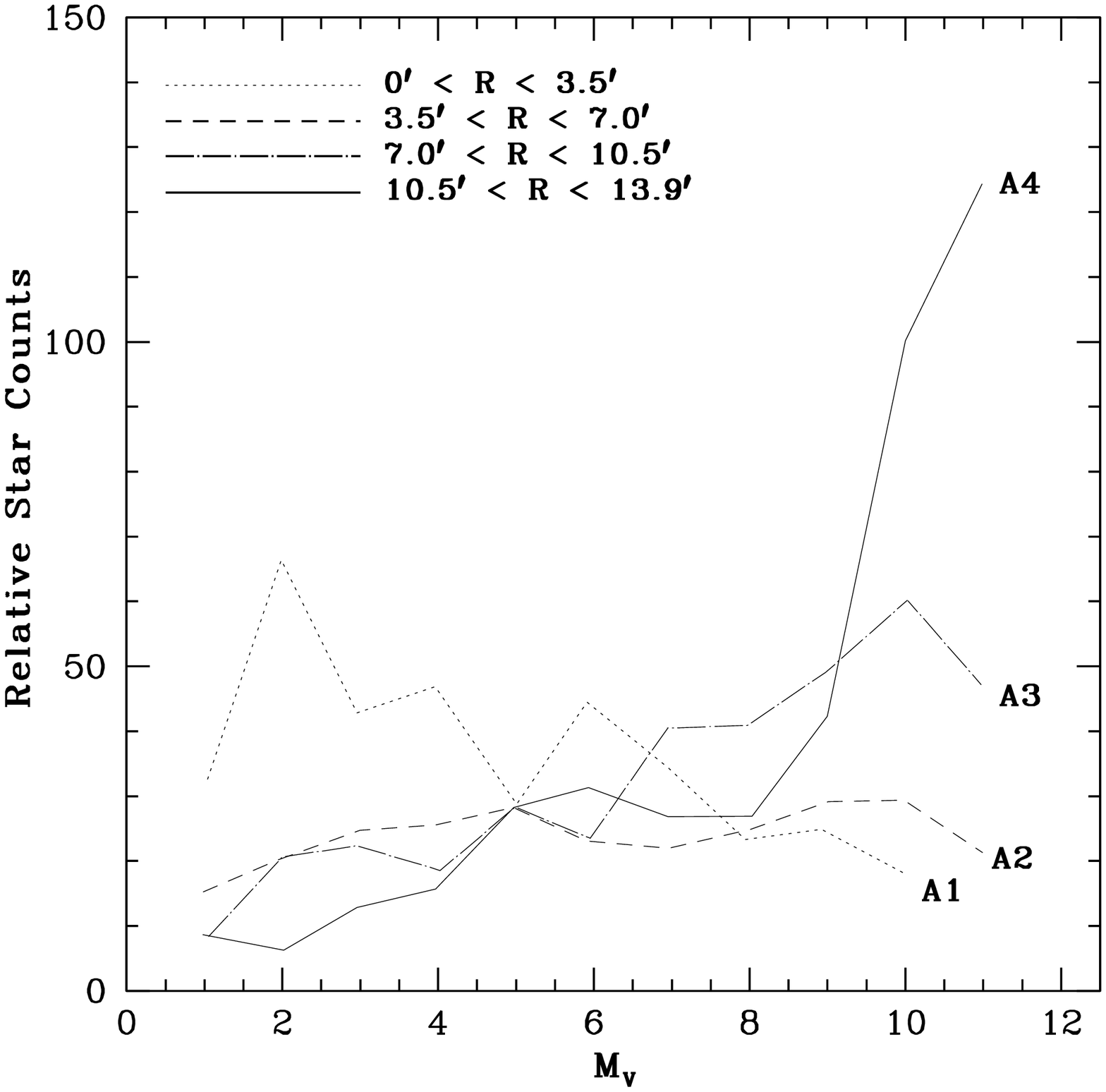]{Luminosity function shown for
four increasing concentric annuli around NGC 2099.  The counts in
all annuli have been normalized to the number at M$_{\rm V}$ = 5.
There is some evidence for mass segregation in the faintest bins.
\label{masssegfig}}

\plotone{Kalirai.fig12.eps}

\clearpage

\figcaption[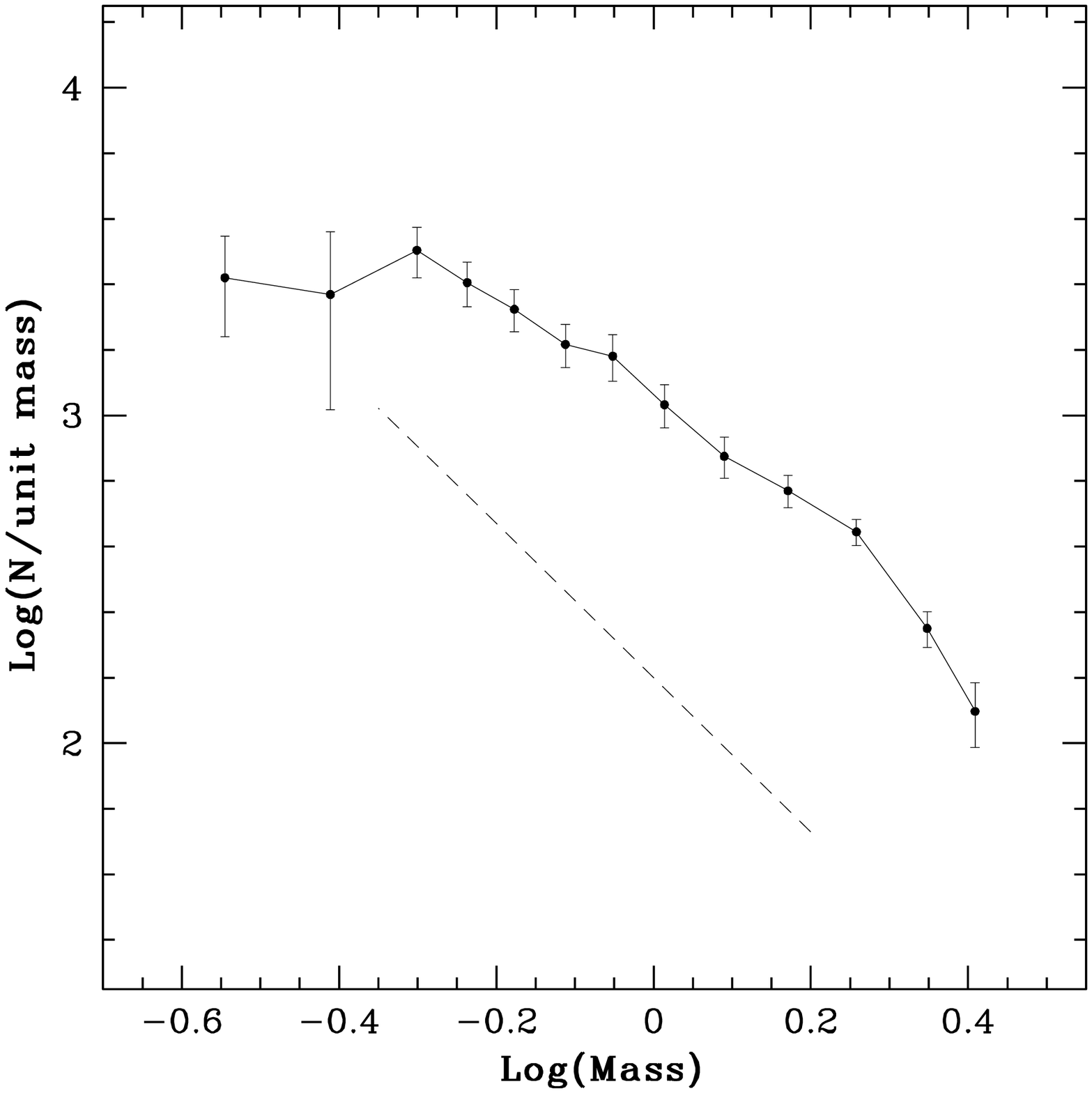]{Global mass function of NGC 2099
(solid, $x$ = 0.60) found to be flatter than a Salpeter IMF
(dashed, $x$ = 1.35) from $\sim$0.5 M$_\odot$ to $\sim$1.8
M$_\odot$.  The slope of the high mass end of the mass function is
found to be very steep, $x$ = 2.5. \label{massfuncfig}}

\plotone{Kalirai.fig13.eps}

\clearpage

\figcaption[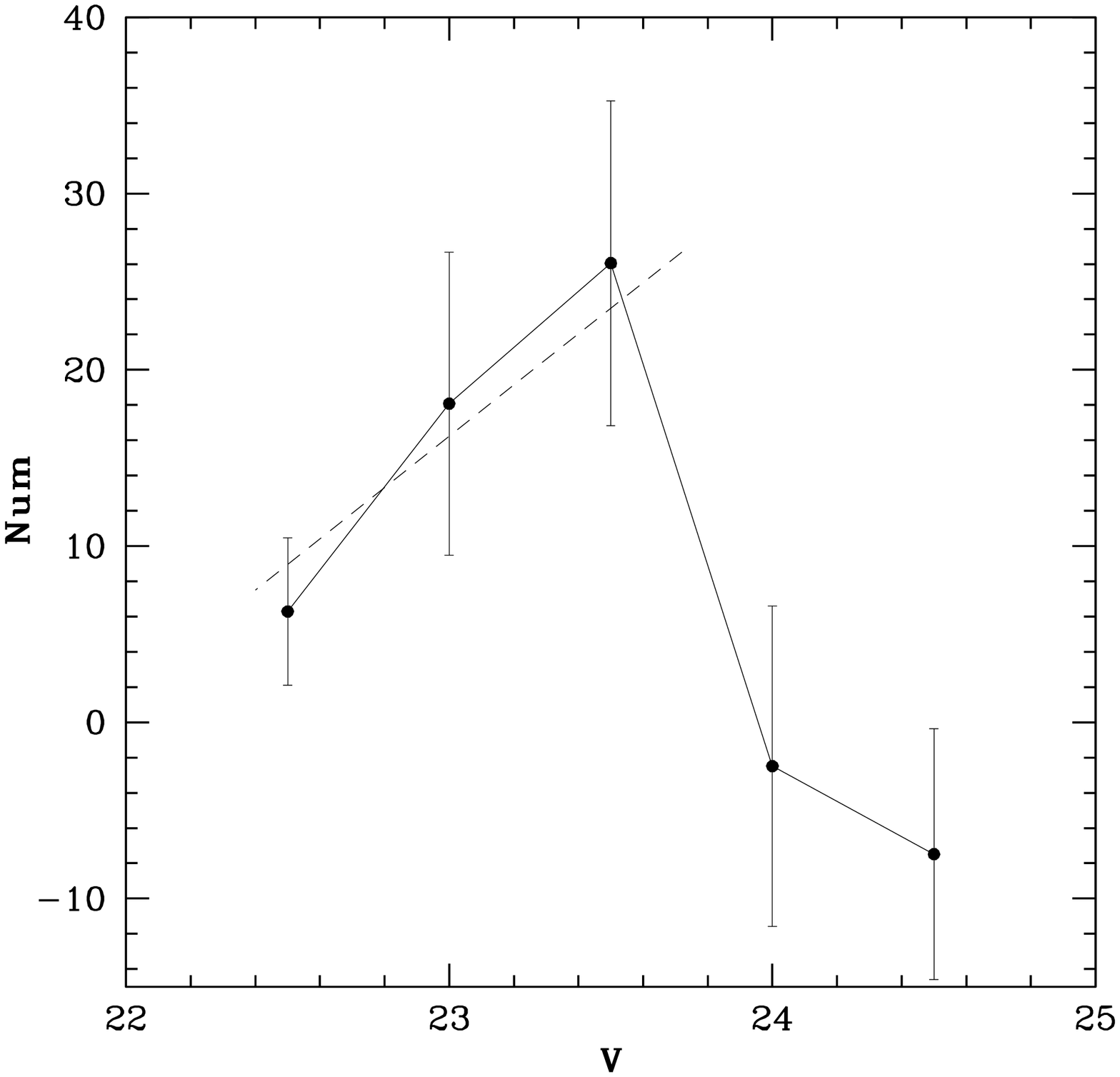]{White dwarf luminosity function
rises to a peak at V = 23.5 (M$_{\rm V}$ = 11.95 $\pm$ 0.30) and
subsequently drops off rapidly (see \S \ref{uncertainty} for error
analysis).  The bright end slope of the luminosity function is in
agreement with theoretical expectations (dashed). The errors bars
include both counting uncertainties and incompleteness errors. For
a 0.70 M$_\odot$ white dwarf cooling sequence, the limiting
magnitude of the cooling white dwarfs provides a white dwarf
cooling age of 566 $\pm \ ^{154}_{176}$ Myrs for NGC 2099.  This
age is in excellent agreement with the main-sequence turn-off age
(520 Myrs) for a core-overshooting model. \label{wdlumfunc}}

\plotone{Kalirai.fig14.eps}

\clearpage

\figcaption[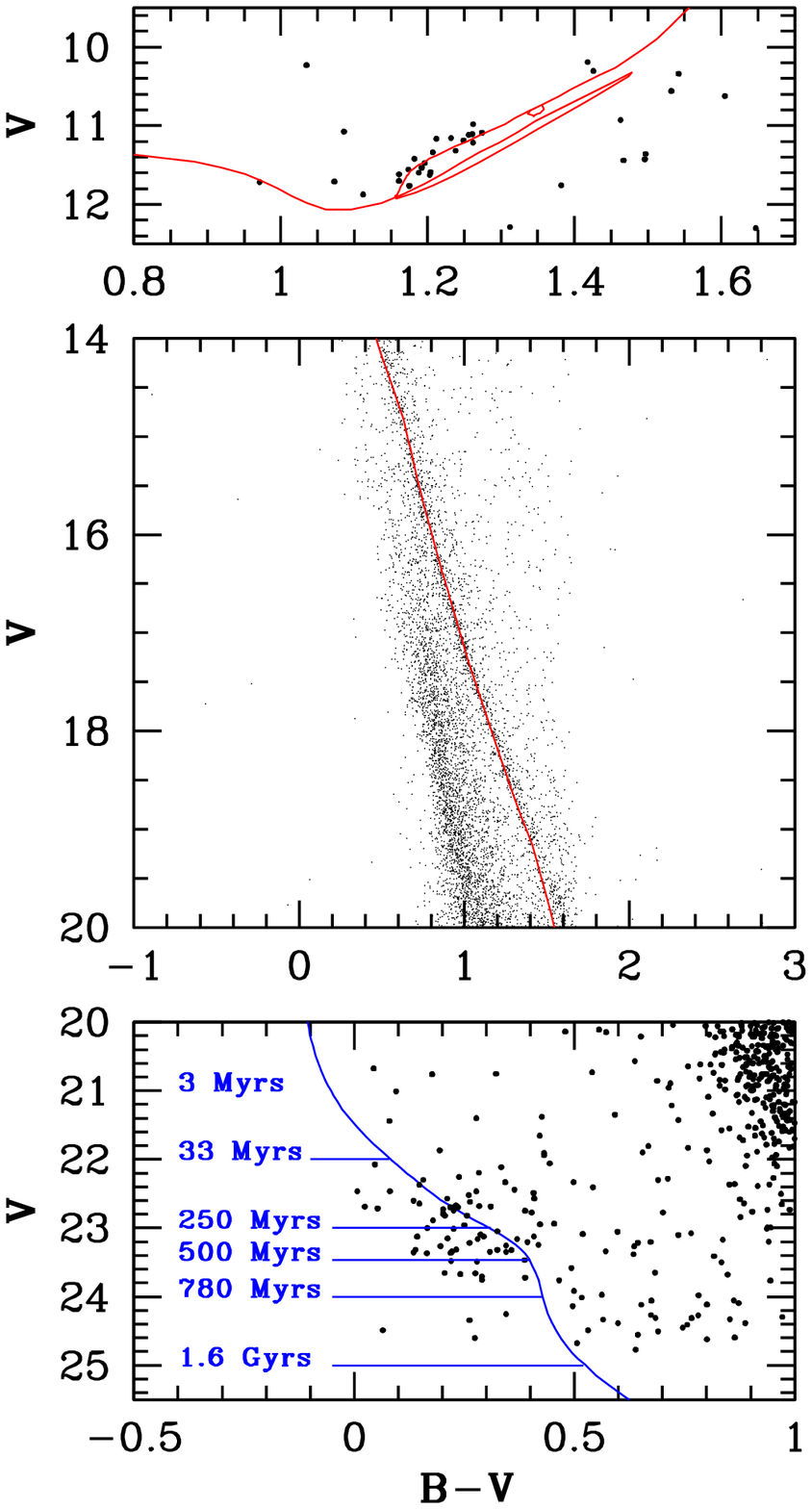]{The theoretical isochrone is in 
excellent agreement with the data for both the red giant clump (top) and 
for the main-sequence from V = 14 to V = 20 (middle).  The white dwarf 
population is clearly evident and is shown with respect to a 0.70 
M$_\odot$ white dwarf cooling sequence and the corresponding cooling ages
(bottom). \label{3plots}}

\plotone{Kalirai.fig15.eps}

\clearpage

%% Tables should be submitted one per page, so put a \clearpage before

%% Two options are available to the author for producing tables:  the
%% deluxetable environment provided by the AASTeX package or the LaTeX
%% table environment.  Use of deluxetable is preferred.
%%

%% Three table samples follow, two marked up in the deluxetable environment,
%% one marked up as a LaTeX table.

%% In this first example, note that the \tabletypesize{}
%% command has been used to reduce the font size of the table.
%% Note also that the \label command needs to be placed

\clearpage

\begin{deluxetable}{ccccc}
\tabletypesize{\scriptsize} \tablecaption{Observational Data for
NGC 2099 \label{table1}} \tablewidth{0pt}
\tablehead{\colhead{Filter} & \colhead{Exposure Time (s)} &
\colhead{No. Images} & \colhead{Seeing ($''$)} &
\colhead{Air-mass}} \startdata

V & 300 & 3 & 0.85 & 1.03 \\

 & 50 & 1 & 0.97 & 1.04 \\

 & 10 & 1 & 0.99 & 1.04 \\

 & 0.5 & 1 & 1.1 & 1.19  \\

\tableline

B & 300 & 3 & 0.79 & 1.03 \\

 & 50 & 1 & 0.85 & 1.03 \\

 & 10 & 1 & 0.85 & 1.03 \\

 & 0.5 & 1 & 1.0 & 1.19   \\

\tableline

R & 50 & 1 & 0.81 & 1.04 \\

  & 10 & 1 & 0.81 & 1.04 \\

\enddata

%% Text for table notes should follow after the \enddata but before
%% the \end{deluxetable}. Make sure there is at least one \tablenotemark
%% in the table for each \tablenotetext.

%% If you use the table environment, please indicate horizontal rules using
%% Do not put multiple tabular environments within a single table.
%% The optional \label should appear inside the \caption command.

\end{deluxetable}

\clearpage

\begin{deluxetable}{cc}
\tabletypesize{\scriptsize} \tablecaption{NGC 2099 Fiducial for 
(m$\rm-$M)$_{\rm V}$ = 11.55, E(B$\rm-$V) = 0.21 \label{table2}} 
\tablewidth{0pt} \tablehead{\colhead{M$_{\rm V}$} & 
\colhead{(B$\rm-$V)$_{\rm o}$}} \startdata
 -0.20  & 0.066 \\
  0.23  & 0.082 \\
  0.70  & 0.082 \\
  1.22  & 0.106 \\
  1.70  & 0.149 \\
  2.20  & 0.219 \\
  2.70  & 0.304 \\
  3.20  & 0.417 \\
  3.74  & 0.485 \\
  4.21  & 0.551 \\
  4.73  & 0.626 \\
  5.21  & 0.711 \\
  5.70  & 0.800 \\
  6.19  & 0.891 \\
  6.71  & 1.011 \\
  7.22  & 1.125 \\
  7.70  & 1.216 \\
  8.19  & 1.321 \\
  8.69  & 1.401 \\ 
  9.22  & 1.462 \\
  9.72  & 1.474 \\
  10.22 & 1.495 \\
  10.71 & 1.517 \\
  11.20 & 1.556 \\
  11.69 & 1.589 \\
  12.12 & 1.600 \\

\enddata

\end{deluxetable}

\clearpage

\begin{deluxetable}{ccccccc}
\tabletypesize{\scriptsize} \tablecaption{Completeness Corrections
\label{table3}} \tablewidth{0pt} \tablehead{\colhead{V mag} &
\colhead{No. Stars Input} & \colhead{No. Stars Recovered
(Cluster/Background)} & \colhead{Completeness Correction
(Cluster/Background)}} \startdata

Main-Sequence \\

11.0-12.0 &34 &34/34 &1/1  \\

12.0-13.0 &24 &24/24 &1/1  \\

13.0-14.0 &26 &26/26 &1/1  \\

14.0-15.0 &46 &46/46 &1/1  \\

15.0-16.0 &34 &34/34 &1/1  \\

16.0-17.0 &54 &54/54 &1/1  \\

17.0-18.0 &50 &48/50 &1.042/1  \\

18.0-19.0 &58 &56/58 &1.036/1  \\

19.0-20.0 &70 &66/68 &1.061/1.029  \\

20.0-21.0 &56 &54/54 &1.037/1.037  \\

21.0-22.0 &100 &94/96 &1.064/1.042  \\

22.0-23.0 &78 &64/66 &1.219/1.182  \\

23.0-24.0 &100 &72/84 &1.389/1.190  \\

24.0-25.0 &62 &36/50 &1.722/1.240  \\

\tableline

White Dwarfs \\

21.0-22.0 &14 &13/14 &1.077/1  \\

22.0-23.0 &30 &26/28 &1.154/1.071  \\

23.0-24.0 &38 &30/32 &1.267/1.188  \\

24.0-25.0 &34 &20/26 &1.700/1.308  \\

\enddata

\end{deluxetable}

\clearpage

\begin{deluxetable}{cccccc}
\tabletypesize{\scriptsize} \tablecaption{Cluster Star Counts
(Raw/Corrected) \label{table4}} \tablewidth{0pt}
\tablehead{\colhead{V mag} & \colhead{A1} & \colhead{A2} &
\colhead{A3} & \colhead{A4} & \colhead{GLOBAL}} \startdata

11.0-12.0 (Raw) &\nodata&\nodata&\nodata&\nodata& 40.1  \\

Corrected &\nodata (\nodata) &\nodata (\nodata) &\nodata (\nodata) &\nodata (\nodata) & 40.1 (7.3) \\

12.0-13.0 (Raw) &32.6&37.8&15.0&11.6&97.0  \\

Corrected &32.6 (5.9)&37.8 (6.8)&15.0 (5.4)&11.6 (5.5)&97.0 (11.9) \\

13.0-14.0 (Raw) &66.4&50.2&36.9&8.3&161.8 \\

Corrected &66.4 (8.5)&50.2 (8.1)&36.9 (7.9)&8.3 (6.6)&161.8 (15.6) \\

14.0-15.0 (Raw) &42.8&61.4&40.1&17.1&159.4 \\

Corrected &42.8 (7.4)&61.4 (9.9)&40.1 (10.1)&17.1 (10.0)&159.4 (18.8) \\

15.0-16.0 (Raw) &46.8&63.5&33.2&20.9&158.5 \\

Corrected &46.8 (8.5)&63.5 (11.9)&33.2 (12.8)&20.9 (14.1)&158.5 (24.0) \\

16.0-17.0 (Raw) &28.7&70.2&50.7&37.6&183.3 \\

Corrected &28.7 (8.2)&70.2 (13.6)&50.7 (15.6)&37.6 (17.3)&183.3 (28.1) \\

17.0-18.0 (Raw) &41.9&52.7&36.6&34.8&155.0 \\

Corrected &44.5 (9.4)&57.4 (13.7)&42.1 (16.0)&41.7 (18.2)&174.2 (31.4) \\

18.0-19.0 (Raw) &32.8&51.3&67.8&31.3&171.1 \\

Corrected &34.4 (8.0)&54.5 (12.0)&72.6 (15.0)&35.7 (15.2)&184.8 (27.1) \\

19.0-20.0 (Raw) &22.6&59.9&71.1&34.7&178.3 \\

Corrected &23.3 (7.4)&61.7 (12.7)&73.3 (15.5)&35.8 (16.0)&183.8 (27.7) \\

20.0-21.0 (Raw) &24.0&69.9&84.8&54.3&208.9 \\

Corrected &24.9 (8.4)&72.5 (14.8)&88.0 (18.4)&56.3 (19.7)&216.7 (34.2) \\

21.0-22.0 (Raw) &16.0&65.8&96.8&119&289.5 \\

Corrected &18.0 (11.0)&73.0 (20.3)&107.9 (26.6)&133.3 (31.4)&323.7 (53.5) \\

22.0-23.0 (Raw) &\nodata&34.6&54.8&115.9&181.7 \\

Corrected &\nodata (\nodata)&52.9 (39.9)&84.6 (58.6)&165.4 (76.9)&277.8 (155.9) \\

23.0-24.0 (Raw) &\nodata&\nodata&\nodata&\nodata& 145.9 \\

Corrected &\nodata (\nodata)&\nodata (\nodata)&\nodata (\nodata)&\nodata (\nodata)& 391.2 (125.6) \\

\enddata

\end{deluxetable}

\clearpage

\begin{deluxetable}{ccc}
\tabletypesize{\scriptsize} \tablecaption{Geometry of Annuli
\label{table5}} \tablewidth{0pt} \tablehead{\colhead{Annulus} &
\colhead{Radius ($'$)} & \colhead{Radius (pixels)}} \startdata

A1 & 0 $\leq$ R $\leq$ 3.5 & 0 $\leq$ R $\leq$ 1019  \\

A2 & 3.5 $\leq$ R $\leq$ 7.0 & 1019 $\leq$ R $\leq$ 2039 \\

A3 & 7.0 $\leq$ R $\leq$ 10.5 & 2039 $\leq$ R $\leq$ 3058 \\

A4 & 10.5 $\leq$ R $\leq$ 13.9 & 3058 $\leq$ R $\leq$ 4049 \\

Global & 0 $\leq$ R $\leq$ 13.9 & 0 $\leq$ R $\leq$ 4049 \\

\enddata

\end{deluxetable}

\clearpage

\begin{deluxetable}{ccc}
\tabletypesize{\scriptsize} \tablecaption{White Dwarf Luminosity
Function \label{table6}} \tablewidth{0pt} \tablehead{\colhead{V mag} &
\colhead{No. Stars} & \colhead{Error}} \startdata

22.5 & 6.3 & 4.2  \\

23.0 & 18.1 & 8.6 \\

23.5 & 26.1 & 9.2 \\

24.0 & -2.5 & 9.1 \\

24.5 & -7.5 & 7.1 \\

\enddata

\end{deluxetable}

\clearpage

%% You can append references to a table using the \tablerefs command.

%% Tables may also be prepared as separate files. See the accompanying
%% sample file table.tex for an example of an external table file.
%% To include an external file in your main document, use the \input
%% command. Uncomment the line below to include table.tex in this
%% sample file.

%\input{table}

%% The following command ends your manuscript. LaTeX will ignore any text


\begin{thebibliography}{}

\bibitem[Adams et al.(2001)]{adams} Adams, J.D.,
Stauffer, J.R., Monet, D.G., Skrutskie, M.F. \& Beichman, C.A. 2001,
\aj, 121, 4, 2053
\bibitem[Barrado y Navascues et. al.(2001)]{barrado} Barrado y Navascues,
D., Stauffer, J.R., Bouvier, J. \& Martin, E.L. 2001, \apj, 546,
Iss. 2, 1006
\bibitem[Becker(1948)]{becker1} Becker, W. 1948, Astr. Nach.,
276, 1
\bibitem[Becker \& Svolopoulos(1976)]{becker2} Becker, W. \&
Svolopoulos, S. 1976, \aaps, 23, 97
\bibitem[Bertelli et al.(1994)]
{bertelli} Bertelli, G., Bressan, A., Chiosi, C., Fagotto, F. \&
Nasi, E. 1994, \aap, 106, 275
\bibitem[Bertin \& Arnouts(1996)]
{bertin1} Bertin, E. \& Arnouts, S. 1996, \aaps, 117, 393
\bibitem[Bessell et al.(1998)]{bessell} Bessell, M.S.,
Castelli, F. \& Plez, B. 1998, \aap, 333, 231
\bibitem[Binney \& Merrifield(1998)]{binney2} Binney, J. \& Merrifield, M. 
1998, Galactic Astronomy, (Princeton: University Press)
\bibitem[Binney \& Tremaine(1987)]{binney} Binney, J. \& Tremaine, S. 1987,
Galactic Dynamics, (Princeton: University Press)
\bibitem[Bolte(1989)]{bolte} Bolte, M. 1989, \apj, 341, 168
\bibitem[de Bruijne, Hoogerwerf \& de Zeeuw(2001)]{deBruijne} de Bruijne,
J.H.J., Hoogerwerf, R. \& de Zeeuw, P.T. 2001, \aap, 367, 111
\bibitem[Canuto \& Mazzitelli(1992)]{canuto2} Canuto, V.M. \& Mazzitelli, 
I. 1992, \apj, 389, 724
\bibitem[Castellani, Degl'Innocenti \& Prada Moroni(2001)]{castellani} 
Castellani, V., Degl'Innocenti, S. \& Prada Moroni, P.G. 2001, \mnras, 320, 1, 66
\bibitem[Clemens(1985)]{clemens} Clemens, D.P. 1985, \apj,
295, 422
\bibitem[Copeland, Jensen \& Jorgensen(1970)]{copeland} Copeland, H.,
Jensen, J.O. \& Jorgensen, H.E. 1970, \aap, 5, 12
\bibitem[Cuillandre(2001)]{cuillandre} Cuillandre, J-C. 2001,
\aap, in preparation
\bibitem[D'Antona(1998)]{dantona} D'Antona, F. 1998, ASP Conference
Series, 142, 157
\bibitem[Francic(1989)]{francic} Francic, S.P. 1989, \aj, 98, 888
\bibitem[Freytag et al.(1996)]{freytag} Freytag, B., Ludwig,
H.G. \& Steffen, M. 1996, \aap, 313, 497
\bibitem[Giebeler(1914)]{giebeler} Giebeler, H. 1914, Bonn Veroff.,
12
\bibitem[Girardi et al.(2000)]{girardi} Girardi, L., Bressan, A., Bertelli, 
G. \& Chiosi, C. 2000, \aaps, 141, 371
\bibitem[Gunn \& Griffin(1979)]{gunn} Gunn, J.E. \& Griffin, R.F. 1979, 
\aj, 84, 753 
\bibitem[Hansen(1999)]{hansen} Hansen, B.M.S. 1999, \apj, 520, 680
\bibitem[Hauschildt, Allard \& Baron(1999)]{hauschildt}
Hauschildt, P.H., Allard, F. \& Baron, E. 1999, \apj, 512, 377
\bibitem[Hawley, Tourtellot \& Reid(1999)]{hawley} Hawley, S.L., 
Tourtellot, J.G. \& Reid, I.N. 1999, \aj, 117, 3, 1341
\bibitem[Hoag et al.(1961)]{hoag} Hoag, A.A.,
Iriarte, B., Johnson, H.L., Hallam, K.L., Mitchell, R.I. \&
Sharpless, S. 1961, Pub. U.S. Naval Observatory, Ser. 2, 17 349
\bibitem[Iben \& Renzini(1983)]{iben} Iben, I. Jr. \& Renzini, A. 
1983, ARAA, 21, 271
\bibitem[Janes, Tilley \& Lynga(1998)]{janes}
Janes, K.A., Tilley, C. \& Lynga, G. 1998, \aj, 95, 771
\bibitem[Jeffreys(1962)]{jeffreys} Jeffreys, W.H., III. 1962, \aj, 67, 532
\bibitem[Kalirai et al.(2001a)]{kalirai1}
Kalirai, J.S., Richer, H.B., Fahlman, G.G., Cuillandre, J.,
Ventura, P, D'Antona, F., Bertin, E., Marconi, G. \& Durrell,
P. 2001a, \aj, 122, 257
\bibitem[Kalirai et al.(2001b)]{kalirai2}
Kalirai, J.S., Richer, H.B., Fahlman, G.G., Cuillandre, J.,
Ventura, P, D'Antona, F., Bertin, E., Marconi, G. \& Durrell,
P. 2001b, \aj, 122, 266
\bibitem[King(1962)]{king}
King, I.R. 1962, \aj, 67, 471
\bibitem[King(1966)]{king2}
King, I.R. 1966, \aj, 71, 276
\bibitem[Koester \& Reimers(1996)]
{reimerskoester8} Koester, D. \& Reimers, D. 1996, \aap, 313, 810
\bibitem[Koester \& Reimers(1993)]
{reimerskoester2} Koester, D. \& Reimers, D. 1993, \aap, 275, 2, 479
\bibitem[Koester \& Reimers(1985)]
{reimerskoester4} Koester, D. \& Reimers, D. 1985, \aap, 153, 260
\bibitem[Koester \& Reimers(1981)]
{reimerskoester6} Koester, D. \& Reimers, D. 1981, \aap, 99, L8
\bibitem[Landolt(1992)]{landolt}
Landolt, A. U. 1992, \apj, 104, 340
\bibitem[Lee, Kang \& Ann(1999)]{lee2}
Lee, S.H., Kang, Y.W. \& Ann, H.B. 1999, JKAS, 14, 2, 61
\bibitem[Lee \& Sung(1995)]
{lee} Lee, S. \& Sung, H. 1995, JKAS, 28, 1, 45
\bibitem[Leggett, Ruiz \& Bergeron(1998)]{leggett}
Leggett, S.K., Ruiz, M.T. \& Bergeron, P. 1998, \apj, 497, 294
\bibitem[Liebert, Dahn \& Monet(1988)]{liebert}
Liebert, J., Dahn, C.C. \& Monet, D.G. 1988, \apj, 332, 891
\bibitem[Lindblad(1954)]{lindblad}
Lindblad, P.O. 1954, St. An., 18, 1
\bibitem[Mermilliod et al.(1996)]{mermilliod}
Mermilliod, J.C., Huestamendia, G., del Rio, G. \& Mayor, M. 
1996, \aap, 307, 80
\bibitem[Nordlund(1909)]{nordlund}
Nordlund, J. 1909, Sv. Ark. Math., 5, 17
\bibitem[O'Brien et al.(1998)]{o'brien} O'Brien, M.S., Vauclair, G., Kawaler, S.D., Watson, T.K., Winget, D.E.,
Nather, R.E., Montgomery, M., Nitta, A., Kleinman, S.J., Sullivan, D.J., Jiang, X.J., Marar, T.M.K., Seetha, S.,
Ashoka, B.N., Bhattacharya, J., Leibowitz, E.M., Hemar, S., Ibbetson, P., Warner, B., van Zyl, L., Moskalik, P.,
Zola, S., Pajdosz, G., Krzesinski, J., Dolez, N., Chevreton, M., Solheim, J-E., Thomassen, T., Kepler, S.O.,
Giovannini, O., Provencal, J L., Wood, M.A. \& Clemens, J.C. 1998, \apj, 495, 458
\bibitem[Perryman et al.(1998)]
{perryman} Perryman, M.A.C., Brown, A.G.A., Lebreton, Y., Gomez,
A., Turon, C., de Strobel, G.C., Mermilliod, J.C., Robichon, N.,
Kovalevsky, J. \& Crifo, F. 1998, \aap, 331, 81
\bibitem[Raboud \& Mermilliod(1998)]
{raboud} Raboud, D. \& Mermilliod, J-C. 1998, \aap, 333, 897
\bibitem[Reid(1992)]{reid}
Reid, N. 1992, MNRAS, 257, 2, 257
\bibitem[Reimers \& Koester(1994)]
{reimerskoester1} Reimers, D. \& Koester, D. 1994, \aap, 285, 451
\bibitem[Reimers \& Koester(1988)]
{reimerskoester3} Reimers, D. \& Koester, D. 1988, \aap, 202, 1, 277
\bibitem[Reimers \& Koester(1988b)]
{reimerskoester7} Reimers, D. \& Koester, D. 1988b, ESO Messenger,
54, 47
\bibitem[Reimers \& Koester(1982)]
{reimerskoester5} Reimers, D. \& Koester, D. 1982, \aap, 116, 2, 341
\bibitem[Renzini \& Fusi-Pecci(1988)]
{renzini} Renzini, A. \& Fusi-Pecci, F. 1988, ARAA, 26, 199
\bibitem[Richer et al.(2000)]
{richer1} Richer, H.B, Hansen, B., Limongi, M., Chieffi, A.,
Straniero, O. \& Fahlman, G.G. 2000, \apj, 529, Issue 1, 318
\bibitem[Richer et al.(1998)]
{richer2} Richer, H.B., Fahlman, G.G., Rosvick, J. \& Ibata, R. 
1998, \apj, 504, L91
\bibitem[Richer et al.(1997)]
{richer3} Richer, H.B., Fahlman, G.G., Ibata, R.A., Pryor, C., Bell,
R.A., Bolte, M., Bond, H.E., Harris, W.E., Hesser, J.E., Holland, S.,
Ivanans, N., Mandushev, G., Stetson, P.B. \& Wood, M.A. 1997, \apj,
484, 741
\bibitem[Sagar \& Griffiths(1998)]{sagar} Sagar, R. \&
Griffiths, W.K. 1998, \mnras, 299, 777
\bibitem[Salpeter(1955)]{salpeter} Salpeter, E.E. 1955, \apj,
121, 161
\bibitem[Saumon \& Jacobson(1999)]{saumon} Saumon, D. \&
Jacobson, S.B. 1999, \apj, 511, L107
\bibitem[Schaller et al.(1992)]{schaller} Schaller, G.,
Shaerer, D., Meynet, G. \& Maeder, A. 1992, \aap, 96, 269
\bibitem[Sung et al.(1999)]{sung} Sung, H., Bessell, M.S., 
Lee, H-W, Kang, Y.H. \& Lee, S-W. 1999, \mnras, 310, 4, 982
\bibitem[Upgren(1966)]{upgren} Upgren, A.R. 1966, \aj,
71, 8, 736
\bibitem[van den Bergh \& Sher(1960)]{vandenberg}
van den Bergh, S. \& Sher, D. 1960, Publ. David Dunlap Obs., 2,
203
\bibitem[Ventura et al.(1998)]{ventura} Ventura, P.,
Zeppieri, A., Mazzitelli, I. \& D'Antona, F. 1998, \aap, 334, 953
\bibitem[von Hippel \& Gilmore(2000)]{vonHippel1} von Hippel, T.
\& Gilmore, G. 2000, \aj, 120, Iss. 3, 1384
\bibitem[von Hippel \& Sarajedini(1998)]{vonHippel2} von
Hippel, T. \& Sarajedini, A. 1998, \aj, 116, Iss. 4, 1789
\bibitem[von Zeipel \& Lindgren(1921)]{vonzeipel} von
Zeipel, H. \& Lindgren, J. 1921, Sv. Vet. H., 61, N15
\bibitem[Weidemann(2000)]{weidemann1} Weidemann, V. 2000, \aap, 363, 647
\bibitem[Weidemann(1987)]{weidemann3} Weidemann, V. 1987, \aap, 188, 74
\bibitem[West(1967)]{west} West, F.R. 1967, \apjs, 14, 359
\bibitem[Wielen(1991)]{wielen} Wielen, R. 1991, ASPCS, 
Formation and Evolution of Star Clusters, 13, 343
\bibitem[Wood(1994)]{wood} Wood, M.A. 1994, AAS Meeting, 185, 4601
\bibitem[Woods \& Fahlman(1997)]{woods} Woods, D. \& Fahlman, G. 1997, \apj,
490, 11
\bibitem[Xiong(1985)]{xiong} Xiong, D.R. 1985, \aap, 150, 133
\bibitem[Zug(1933)]{zug} Zug, R.S. 1933, Lick Obs. Bull., 16, 130



\end{thebibliography}
\end{document}